%% file: CO_c3po.tex
\documentclass[twocolumn]{aastex701}
\usepackage{apjfonts} 

\newif\preprint 

\usepackage{amsmath,amstext}
\usepackage[T1]{fontenc}
\usepackage{comment}
\usepackage{natbib}
\usepackage{multirow}
\usepackage{graphicx}
\usepackage{tabularx}
\usepackage{xcolor}
\usepackage{array}
\usepackage{subcaption}
\usepackage{verbatim}
\usepackage{hyperref}
\hypersetup{
    colorlinks=true,
    linkcolor=blue,
    filecolor=magenta,      
    urlcolor=cyan,
  }
  \usepackage{footmisc}
\graphicspath{{./}{figures/}}

\begin{document}
\include{defs}

\newcommand{\todo}[1]{\textcolor{aggiemaroon}{\tt #1}}

\title{\large \bf Are AGN Required? Updated Photoionization Models for Rest-UV Line Diagnostics in Reionization-era Galaxies}

\author[0000-0001-7593-9205]{Jiayang Yang}
\affiliation{Department of Physics and Astronomy, Texas A\&M University, College Station, TX, 77843-4242 USA}
\affiliation{George P.\ and Cynthia Woods Mitchell Institute for Fundamental Physics and Astronomy, Texas A\&M University, College Station, TX, 77843-4242 USA}
\email{annabellayang@tamu.edu}

\author[0000-0001-7503-8482]{Casey Papovich}
\affiliation{Department of Physics and Astronomy, Texas A\&M University, College Station, TX, 77843-4242 USA}
\affiliation{George P.\ and Cynthia Woods Mitchell Institute for Fundamental Physics and Astronomy, Texas A\&M University, College Station, TX, 77843-4242 USA}
\email{papovich@tamu.edu}

\author[0000-0003-3424-3230]{Weida Hu}
\affiliation{Key Laboratory for Research in Galaxies and Cosmology, Shanghai Astronomical Observatory, Chinese Academy of Sciences, 80 Nandan Road, Shanghai 200030, China}
\email{weidahu.astro@gmail.com}

\author[0000-0001-6251-4988]{Taylor A. Hutchison}
\affiliation{Astrophysics Science Division, Code 660, NASA Goddard Space Flight Center, 8800 Greenbelt Rd., Greenbelt, MD 20771, USA}
\affiliation{Department of Astronomy, University of Maryland, Baltimore County, MD 21250, USA}
\affiliation{Center for Research and Exploration in Space Science and Technology, NASA/GSFC, Greenbelt, MD 20771 USA}
\affiliation{Department of Astronomy, The University of Texas at Austin, Austin, TX 78712, USA}
\affiliation{Cosmic Frontier Center, The University of Texas at Austin, Austin, TX 78712, USA}
\email{}

\author[0000-0003-2366-8858]{Rebecca L. Larson}
\altaffiliation{Giacconi Postdoctoral Fellow}
\affil{Space Telescope Science Institute, 3700 San Martin Drive, Baltimore, MD 21218, USA}
\email{rlarson@stsci.edu}

\author[0000-0000-0000-0000]{Pablo Arrabal Haro}
\altaffiliation{NASA Postdoctoral Fellow}
\affiliation{Astrophysics Science Division, NASA Goddard Space Flight Center, 8800 Greenbelt Rd, Greenbelt, MD 20771, USA}
\email{pablo.arrabalharo@nasa.gov}

\author[0000-0003-1282-7454]{Anthony J. Taylor}
\affiliation{Department of Astronomy, The University of Texas at Austin, Austin, TX 78712, USA}
\affiliation{Cosmic Frontier Center, The University of Texas at Austin, Austin, TX 78712, USA}
\email{anthony.taylor@austin.utexas.edu}

\author[0000-0002-9426-7456,gname=Ray,sname=Garner,suffix=III]{Ray Garner, III}
\affiliation{Department of Physics and Astronomy, Texas A\&M University, College Station, TX, 77843-4242 USA}
\affiliation{George P.\ and Cynthia Woods Mitchell Institute for Fundamental Physics and Astronomy, Texas A\&M University, College Station, TX, 77843-4242 USA}
\email{ray.three.garner@gmail.com}

\author[0000-0001-5749-5452]{Kaila Ronayne}
\affiliation{Department of Physics and Astronomy, Texas A\&M University, College Station, TX, 77843-4242 USA}
\affiliation{George P.\ and Cynthia Woods Mitchell Institute for Fundamental Physics and Astronomy, Texas A\&M University, College Station, TX, 77843-4242 USA}
\email{kaila\_ronayne@tamu.edu}

\author[0000-0001-7151-009X]{Nikko J. Cleri}
\affiliation{Department of Astronomy and Astrophysics, The Pennsylvania State University, University Park, PA 16802, USA}
\affiliation{Institute for Computational and Data Sciences, The Pennsylvania State University, University Park, PA 16802, USA}
\affiliation{Institute for Gravitation and the Cosmos, The Pennsylvania State University, University Park, PA 16802, USA}
\email{cleri@psu.edu}

\author[0000-0001-8519-1130]{Steven L. Finkelstein}
\affiliation{Department of Astronomy, The University of Texas at Austin, Austin, TX 78712, USA}
\affiliation{Cosmic Frontier Center, The University of Texas at Austin, Austin, TX 78712, USA}
\email{stevenf@astro.as.utexas.edu}

\author[0000-0001-8534-7502]{Bren E. Backhaus}
\affiliation{Department of Physics and Astronomy, University of Kansas, Lawrence, KS 66045, USA}
\email{bren.backhaus@ku.edu}

\author[0000-0002-9921-9218]{Micaela B. Bagley}
\affiliation{Southeastern Universities Research Association, Washington, DC 20005, USA}
\affiliation{Astrophysics Science Division, NASA Goddard Space Flight Center, Greenbelt, MD 20771, USA}
\affil{Department of Astronomy, The University of Texas at Austin, Austin, TX, USA}
\email{micaela.b.bagley@nasa.gov}

\author[0000-0002-6348-1900]{Justin W. Cole}
\altaffiliation{NASA FINESST Investigator}
\affiliation{Department of Physics and Astronomy, Texas A\&M University, College Station, TX, 77843-4242 USA}
\affiliation{George P.\ and Cynthia Woods Mitchell Institute for Fundamental Physics and Astronomy, Texas A\&M University, College Station, TX, 77843-4242 USA}\
\email{jwc68@tamu.edu}

\author[0000-0001-9187-3605]{Jeyhan S. Kartaltepe}
\affiliation{Laboratory for Multiwavelength Astrophysics, School of Physics and Astronomy, Rochester Institute of Technology, 84 Lomb Memorial Drive, Rochester, NY 14623, USA}
\email{jeyhan@astro.rit.edu}

\begin{abstract}
Rest-frame ultraviolet (UV) emission lines in galaxies provide important probes of the ionizing radiation, chemical abundances, and interstellar medium conditions. Here, we consider UV emission-line diagnostics for galaxies with the observed physical properties of young, metal-poor stellar populations in the epoch of reionization.  We study eight $z \sim 7$--$9$ galaxies using deep JWST/NIRSpec G140M and G395M spectroscopy from the Carbon-3 plus Oxygen survey. We measure rest-frame UV and optical emission lines, \civ\ $\lambda\lambda1548,1551$, \heii\ $\lambda1640$, \ion{O}{3}] $\lambda1666$, \ciii\ $\lambda\lambda1907,1909$, [O\,{\sc ii}] $\lambda\lambda3727,3730$, H$\gamma$, [O\,{\sc iii}] $\lambda4364$, H$\beta$, and [O\,{\sc iii}] $\lambda\lambda4960,5008$. The galaxies have low oxygen abundances, $12+\log(\mathrm{O/H})=7.50$--$7.90$, low C/O ratios, $\mathrm{C/O}=0.10$--$0.32$, and high electron densities, $n_e$(\ciii)$=(1.9$--$4.9)\times10^4\,\mathrm{cm}^{-3}$. We compute \textsc{Cloudy} photoionization models using \textsc{BPASS} and BC03 stellar populations to test how lower C/O and higher gas density affect the UV diagnostics. We find that reducing C/O and increasing the gas density shift the star-forming grids into regions conventionally classified as composite or AGN-like. We therefore conclude that positions within these regions alone do not establish the presence of an AGN in the galaxies studied here.  However, some lines remain enigmatic, including strong \heii, which is not reproduced by the updated models.  These results highlight the need for UV diagnostic boundaries that account for the abundance patterns and gas conditions of galaxies during the epoch of reionization.

\end{abstract}


\section{Introduction}
\label{sec:intro}
The Epoch of Reionization (EoR) was the last global phase transition of the intergalactic medium (IGM) of the Universe, during which the ionizing photons produced by the first generations of luminous star-forming galaxies and active galactic nuclei (AGN) were emitted into their surrounding IGM. Measurements of the cosmic microwave background (CMB) provide a constraint on the midpoint of reionization, which occurred at $z \sim 7$–8 \citep{Planck_Collaboration2020}. The prevailing view is that the bulk of the ionizing photon budget was supplied predominantly by low-mass, star-forming galaxies \citep[e.g.,][]{Wise2014, Finkelstein2019, Jung2025}. However, JWST observations have revealed a population of accreting black holes at this epoch \citep{Harikane2023, Akins2024, Juodzbalis2024}, renewing debate over the contribution of AGN to the ionizing background and motivating models in which they dominate hydrogen reionization \citep{Madau2024}. This complicates the interpretation of the earliest galaxies' spectra. Distinguishing star formation from AGN, characterizing the chemical abundance pattern, and understanding the gas-phase conditions of EoR galaxies are therefore essential to understanding how the Universe was reionized.

The emission properties of galaxies and AGN during the EoR are also crucially important. Ratios of collisionally excited and hydrogen recombination lines constrain gas-phase metallicity, while auroral-to-nebular line ratios and resolved doublets probe electron temperature and density, respectively \citep[e.g.,][]{Kewley2002, Izotov2006, Steidel2016, Nicholls2017}. At $z\gtrsim 6$, many of the key rest-frame UV and optical emission-line diagnostics are redshifted into the near-infrared wavelength range accessible to JWST/NIRSpec \citep{Jakobsen2022}. NIRSpec provides spectroscopic coverage of approximately $0.6$--$5.3\,\mu$m, which has enabled measurements of faint rest-UV and temperature-sensitive auroral emission lines in galaxies during the EoR \citep[e.g.,][]{ArellanoCordova2022, Curti2023, Finkelstein2025}.

Among the most powerful diagnostics accessible in these spectra are the rest-UV emission lines \civ\, $\lambda\lambda$1548,1551, \heii\, $\lambda$1640, \oiii]\, $\lambda$1666, and \ciii\, $\lambda\lambda$1907,1909. 
These lines probe a wide range of ionization energies: \ciii\, traces C$^{2+}$ with an ionization potential of 24.4--47.9 eV; O {\sc iii}] traces O$^{2+}$ over 35.1--54.9 eV; \civ\, traces C$^{3+}$ over 47.9--64.5 eV; and the \heii\, recombination emission requires a photon energy above 54.4 eV. Thus, their relative strength of these lines constrains the hardness of the ionizing spectrum. Together with the rest-frame optical emission lines [O\,{\sc ii}] $\lambda\lambda3727,3730$, H$\gamma$, [O\,{\sc iii}] $\lambda4364$, H$\beta$, and [O {\sc iii}] $\lambda\lambda4960,5008$, these measurements constrain the gas-phase metallicity, C/O ratio, ionization state, electron density, and the electron temperature of the ionized gas. Before JWST, the studies of these rest-UV diagnostics in star-forming galaxies relied primarily on nearby analogs and galaxies at intermediate redshift \citep[e.g.,][]{Stark2014, Stark2016, Maseda2017, Mingozzi2022}. With JWST observations, we can now detect these UV and optical diagnostics in individual reionization-era galaxies.

Emission-line diagnostics are also useful for distinguishing star-forming galaxies from AGN. Multiple studies have used photoionization models to determine regions based on emission-line ratios that separate star-forming galaxies from AGN, and to identify ``composite'' regions that contain both star-forming and AGN models \citep{Hirschmann2017, Hirschmann2019, Hirschmann2023}. Many commonly used models use ionization and nebular properties of galaxies in the local Universe or at moderate redshift ($z\sim 1-3$) \citep{Mingozzi2022, Mingozzi2024}, so their abundance patterns and densities restrict the range of predicted emission line ratios. These models often assume abundances scale with Solar, including a solar C/O (0.55 according to \citet{Asplund2009}, 0.44 assumed in \citet{Gutkin2016}, \citet{Feltre2016}, and \citet{Plat2026}, and 0.36 assumed in \textsc{Bagpipes} \citep{Carnall2018}) and a fiducial gas density $n_H = 100~\mathrm{cm^{-3}}$. Such a fiducial density is broadly consistent with measurements of low-ionization gas in local star-forming galaxies \citep[e.g.,][]{Mingozzi2022}. However, C/O varies with increasing metallicity even locally, with substantially subsolar values measured in metal-poor \hii\, regions \citep[e.g.,][]{Garnett1999, Berg2019b}. These fiducial abundance patterns and densities may therefore not represent the physical conditions of galaxies in the EoR, where subsolar C/O ratios \citep[e.g.,][]{Jones2023, Hu2024} and elevated electron densities \citep[e.g.,][]{Isobe2023, Hu2024} have been measured.

One challenge to the interpretation of high redshift galaxies observations is that evolving nebular gas properties impact predicted emission-line ratios \citep{Cleri2025, Kewley2013}.  For example, EoR galaxies are predicted and observed to have sub-solar C/O ratios. Oxygen is mainly produced by massive stars and released immediately through core-collapse supernovae (CCSN), while carbon comes from both massive stars and intermediate-mass stars, with a delayed enrichment history. Previous research shows a trend of increasing C/O with metallicity in dwarf galaxies and metal-poor stars, and low-metallicity systems have low C/O, albeit with significant scatter \citep{Garnett1995, Akerman2004, Berg2016, Berg2019b}. Galaxies observed less than $\sim$1~Gyr after the Big Bang (i.e., $z \gtrsim 5$) have insufficient time to build up to solar C/O, and recent studies confirm sub-solar C/O in $z> 6$ systems \citep{ArellanoCordova2022, Jones2023, Bunker2023, Hu2024}. A lower C/O ratio impacts emission-line predictions, as it will, for example, lower the strength of \ciii\ and \civ\ relative to  \oiii] and \heii.

In addition, observations of EoR galaxies find that compared to lower-redshift galaxies, they are spatially more compact, more intensely star-forming (i.e., with higher specific SFRs, sSFRs), and they have higher gas densities. Higher ionization parameters may also be associated with more massive young star clusters at early times \citep{Scharre2026}. JWST measurements of density-sensitive doublets, e.g.,\oii\ $\lambda \lambda 3727,3730$ and \sii\ $\lambda \lambda 6718,6733$, find electron densities well above the local average, with $n_H$ commonly greater than $\sim$$10^{3}~\mathrm{cm^{-3}}$ in younger systems \citep{Isobe2023, Topping2025}. Moreover, the \ciii\ $\lambda \lambda 1907,1909$ doublet, which traces the intermediate ionization zone C$^{2+}$ at 24.4--47.9 eV, shows a similar trend of increasing $n_e$(\ciii) with redshift \citep{Mason2023, Topping2025}, but with densities approaching $10^4$ cm$^{-3}$ or higher. Since the emissivities of collisionally excited UV lines and the ionization structure of the nebula both depend on the density, adopting a single low fiducial density can systematically bias the line ratios that standard grids predict \citep{Martinez2026}. 

Given the observations of high-ionization rest-UV emission lines of EoR galaxies, and given the galaxies' physical properties, it is prudent to reconsider the parameter space of emission-line ratios used to characterize the ionizing source.   Previous work interpreted strong emission of \civ\ and elevated \civ/\ciii\ as potential evidence for AGN activity \citep[e.g.,][]{Matsuoka2009,Feltre2016, Mascia2023, Hirschmann2023}. However, this line ratio and line strength can be indicative of hard radiation fields produced by young, metal-poor stellar populations, including the effects of binary evolution, and higher nebular gas densities \citep{Stark2014, Berg2018, Senchyna2023, Lecroq2024}. Radiative shocks can also overlap some of these diagnostic regions \citep{Hirschmann2019}. The interpretation of the emission line ratios and line strengths depends on the assumed conditions of the ISM in the models. 

In this paper, we test directly whether stellar photoionization models with lower C/O ratios and higher electron densities, which may be more representative of the EoR environment, can reproduce the rest-UV emission-line ratios measured in EoR galaxies at $7 \lesssim z \lesssim 9$ from the Carbon 3 Plus Oxygen (C3PO) survey without requiring an AGN. We compute nebular grids with Cloudy\,c25.00 \citep{Gunasekera2025-CLOUDY}, using \textsc{BPASS} stellar populations that include binary stars \citep{Eldridge2017, Stanway2018} and the \citet{BruzualCharlot2003}  models.  We re-examine the star-forming-AGN classifications derived using rest-UV emission lines \citep[e.g.,][]{Jaskot2016, Gutkin2016, Hirschmann2023}, where we systematically reduce C/O and increase $n_H$ over the range of parameter space motivated by JWST observations.  We then compare the new model predictions with the emission lines measured from our galaxies, and compare them with direct constraints on the gas electron densities and C/O and O/H abundances derived from the deep C3PO NIRSpec G140M and G395M spectra. Our purpose is not to argue that AGN are absent from the reionization era, but rather to determine whether the observed AGN-like UV line ratios require an AGN once the stellar photoionization models account for the nebular conditions in these galaxies. 

This paper is organized as follows. Section~\ref{sec: data} describes the C3PO observations, the data reduction, the sample selection, and the spectroscopic data from the C3PO survey. Section~\ref{sec: characteristic} presents the emission-line measurements and the derived physical properties. Section~\ref{sec:models} describes the construction of our CLOUDY AGN and Star-Forming models and the explored parameter spaces. Section~\ref{sec:results} compares these models with line ratio measurements with selected JWST galaxies and their local analogs. Section~\ref{sec:discussion} discusses the implications for Star-Forming/AGN identification at high redshift. Section~\ref{sec:conclusion} summarizes our main conclusions. Throughout this work, we assume a flat $\Lambda$CDM cosmology with $H_0 = 70$~km~s$^{-1}$~Mpc$^{-1}$, $\Omega_m = 0.3$, and $\Omega_\Lambda = 0.7$. We denote \ciii\ for the sum of \ciii\ $\lambda\lambda$1907,1909 doublet, \civ\ for sum of the \civ\ $\lambda\lambda$1548,1551 doublet, and O\,{\sc iii]} for the sum of UV O\,{\sc iii]}\ $\lambda\lambda$1661,1666 doublet. All emission-line wavelengths are given in vacuum. We quote abundances relative to the solar values of \citet{Asplund2009}.

\section{Data}
\label{sec: data}

\subsection{C3PO Survey}
\label{sec:c3po}
The galaxies analyzed throughout this paper were observed as part of the C3PO program \citep[JWST GO-5943; PIs: C. Papovich, W. Hu, and T. Hutchison; MAST DOI: 10.17909/t83k-jc86;][]{Papovich2026, Hu2026}. The C3PO survey was designed to be able to use rest-frame UV and optical emission-line diagnostics on galaxies at EoR to characterize their physical conditions. These include the ionization states, electron temperatures, oxygen and carbon abundances, stellar populations, star formation rates, electron densities, dust attenuation, and dust laws. In this work, we primarily focus on the oxygen and carbon abundances, and electron densities.

The C3PO survey targeted galaxies with $z = 6.9$--$8.9$ across two NIRSpec MSA pointings and obtained deep spectroscopy with the G140M/F100LP (13.3~hrs) and G395M/F290LP (10.3~hrs) configurations. These configurations provide coverage of the key rest-frame UV and optical emission features used throughout this work \citep[][]{Papovich2026, Hu2026}.

The C3PO targets have extensive existing JWST observations that constrain data selections. These prior observations include the  Cosmic Evolution Early Release Science Survey (CEERS; ERS-1345; PI: S. Finkelstein; MAST DOI: 10.17909/z7p0-8481; \citealt{Finkelstein2025}), the Red Unknowns: Bright Infrared Extragalactic Survey (RUBIES; GO-4233; PIs: A. de Graaff and G. Brammer; \citealt{deGraaff2025}), and the CANDELS-Area Prism Epoch of Reionization Survey (CAPERS; GO-6368; PI: M. Dickinson; MAST DOI: 10.17909/0q3p-sp24; \citealt{Dickinson_2024}).   The dataset has been used to study the properties of a Ly$\alpha$-emitting galaxy at $z\sim 9$ \citep{Hu2026} and the UV spectra of LRDs at $z \sim 7-8$ \citep{Papovich2026}. 

\subsection{Data Reduction and Calibration}
The primary C3PO NIRSpec data reduction is discussed in \cite{Papovich2026}.  Briefly, the data were reduced with a custom pipeline built on the JWST Calibration Pipeline v1.20.2 \citep{Bushouse2025}, using the \texttt{1481.pmap} Calibration Reference Data System (CRDS) context.  In addition, for this work, we apply an additional correction for residual 1/f noise between the \texttt{detector1} and \texttt{spec2} stages, and we apply the ``picture frame'' artifact (H. Akins, in prep.).  We then  extract final 1D spectra via S/N-weighted optimal extraction \citep{horne86}.

We also experimented with extending the usable wavelength range of G140M from 0.97--1.89~$\mu$m to 0.97--3.2~$\mu$m by modifying the flat-field reference files, exploiting NIRSpec's long-pass filters that pass light beyond each grating's nominal red cutoff. These ranges are normally excluded to avoid second-order contamination, which is negligible for our line-based analysis except in the case of an exact factor-of-two wavelength coincidence between strong lines---not present in our sample.

To calibrate the extended wavelength range, we used deep observations (29.2 hours G140M, 7.9 hours G235M) from the SPectroscopic Ultra-deep Reionization-era Survey (SPURS; GO-9214; PIs: C. Mason and D. Stark; \citep{Chen2026}). We first extended the flat-fields by holding their last valid value constant to the new maximum wavelength, then re-reduced SPURS to obtain spectra calibrated in their native ranges but uncalibrated in the extensions. For lines falling in both the extended range of G140M and the calibrated range of G235M (e.g., H$\alpha$ at $z=3$, observed at 2.62~$\mu$m), we compared fluxes to derive per-wavelength correction factors. We then fit quadratics to these correction factors across the entire dataset, and used them to recalibrate the extended flat-fields. Re-reducing SPURS with these final files and repeating the cross-grating comparison shows $\sim$10\% (1$\sigma$) residual scatter with no redshift dependence, indicating $\sim$10\% flux calibration accuracy in the extended ranges. These final calibrated spectra form our dataset.

All reduced data, the reduction pipeline, and the extended flat-field files are hosted on CAMPFIRE (COSMOS Archive of MultiPle-Field Internal Reductions \& Extractions), a platform for accessing reduced JWST imaging and spectroscopic data primarily in the COSMOS field (H. Akins, in prep.).

\subsection{Sample Selection}
\begin{figure*}[!ht]
    \centering
    \includegraphics[width=1\linewidth]{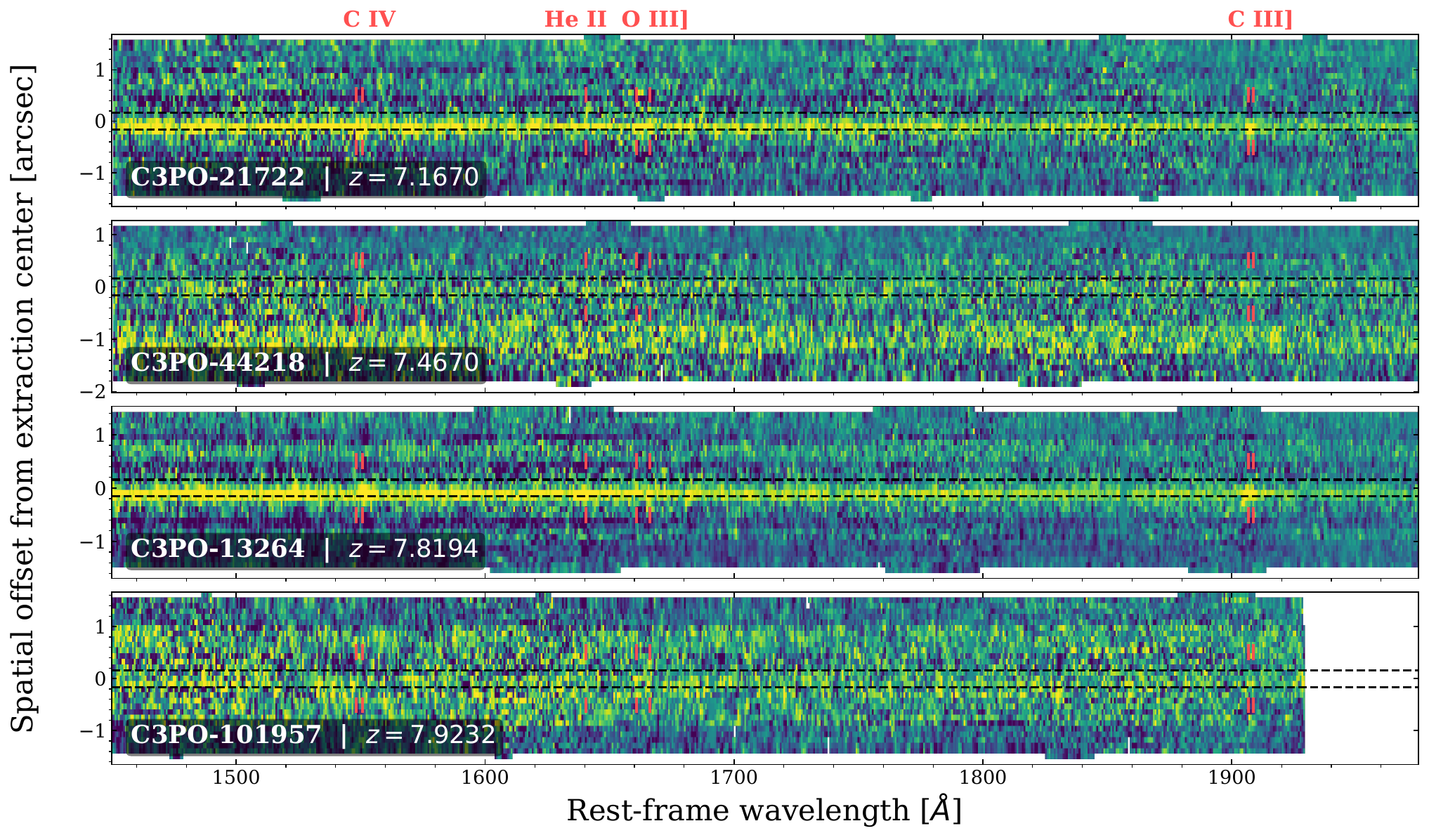}
    \caption{The 2D rest-frame JWST/NIRSpec G140M spectra of example C3PO targets included in this work. The vertical lines mark the expected location of the diagnostic emission lines: $\civ\,  \lambda \lambda1548,1551,\, \heii\, \lambda1640,\, \oiii]\, \lambda \lambda 1661,1666$, and $\ciii\, \lambda \lambda 1907,1909$. The horizontal lines mark the extraction window for the 1d spectra.}
    \label{fig: data_2d}
\end{figure*}

\begin{figure*}[!ht]
    \centering
    \includegraphics[width=1\linewidth]{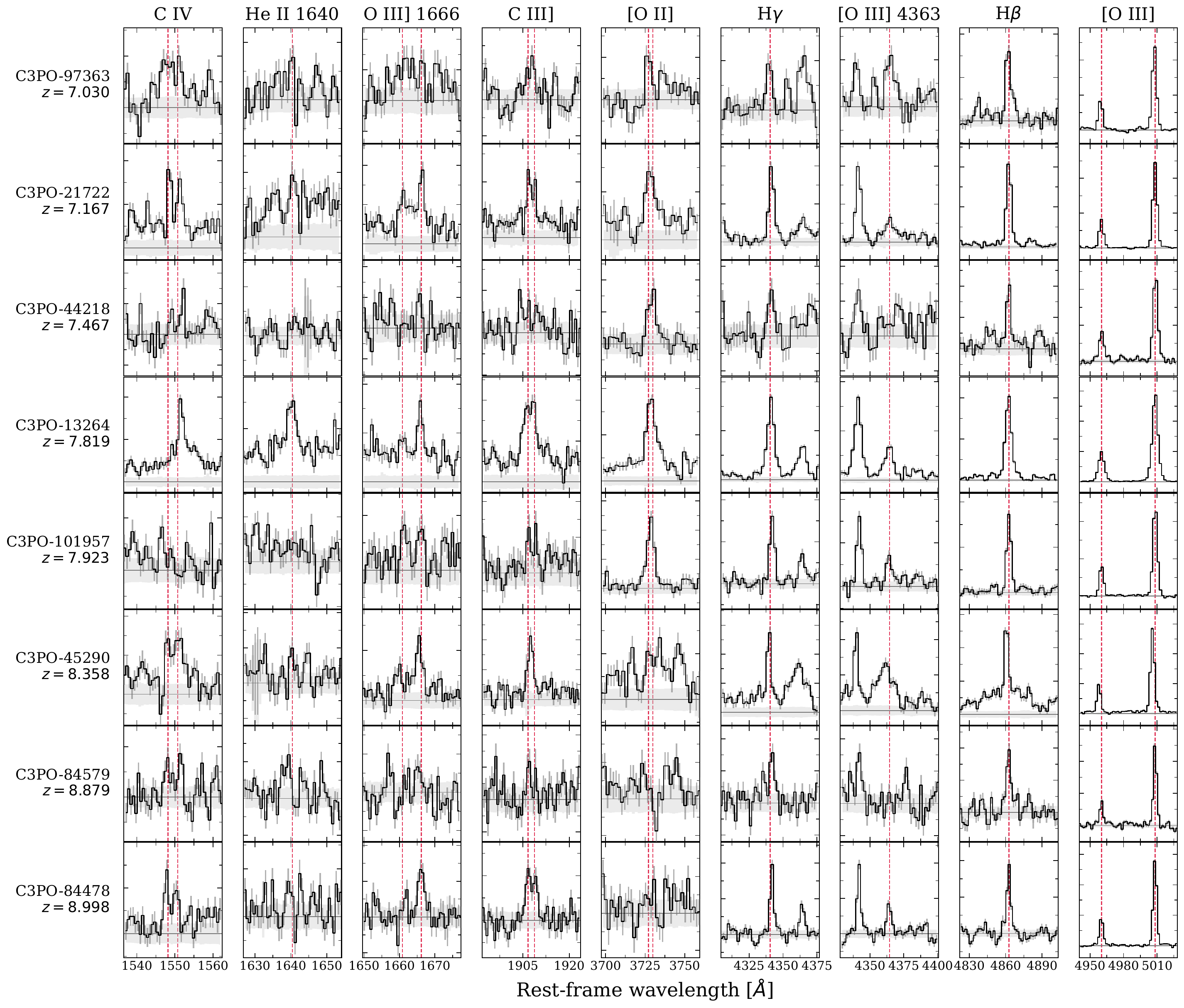}
    \caption{Rest-frame 1D spectral windows centered on the main UV and optical emission lines for the C3PO sample used throughout the paper. Each row corresponds to an individual galaxy. The columns show the spectral regions around \civ\ $\lambda\lambda1548,1551$, \heii\ $\lambda1640$, O,{\sc iii]}\ $\lambda\lambda1661,1666$, \ciii\ $\lambda\lambda1907,1909$, [O {\sc ii]} $\lambda\lambda3727,3730$, \hg, [O\,{\sc iii}]$\lambda4364$, \hb, and O\,{\sc iii}\, $\lambda\lambda4960,5008$ from left to right, respectively. Except for the final O\,{\sc iii}\, $\lambda\lambda4960,5008$ panel, all spectral windows span a common velocity range of $\sim5000~{\rm km\,s^{-1}}$ around the central wavelength of each line or doublet. The spectra are shown in the rest frame using the systemic redshifts listed at left. The vertical axis shows the observed flux density, $F_\nu$ (Jy), plotted against rest-frame wavelength without rescaling the flux density. Each panel has an independent vertical scale. The black lines show the extracted 1D spectra, and the gray shaded regions indicate the corresponding $1\sigma$ uncertainty. Vertical red dashed lines mark the expected rest-frame wavelengths of each emission line.}
    \label{fig: data_1d}
\end{figure*}

The final calibrated spectra provide complementary rest-frame UV and optical coverage through the G140M/F100LP and G395M/F290LP configurations. The G140M grating has a spectral resolving power of $R \sim 1000$ over a wavelength range of approximately $0.97$–$1.89\,\mu{\rm m}$, corresponding to rest-frame wavelengths of $\sim 1200$–$2350$\,\AA\ at $z \sim 7$. For the primary redshift range, $6.9 < z < 8.9$, this spectral coverage includes the principal rest-frame ultraviolet diagnostic emission lines examined in this work: \civ\ $\lambda\lambda 1548,1551$, \heii\ $\lambda 1640$, O\,{\sc iii}]\ $\lambda \lambda 1661, 1666$, and \ciii\ $\lambda\lambda 1907,1909$. In addition, we make use of observations obtained with the G395M grating, which cover the rest-frame optical emission lines \oii\ $\lambda\lambda 3727,3730$, [\oiii]\ $\lambda 4364$, \oiii\ $\lambda\lambda 4960,5008$, \hg, and \hb. We use these lines to derive the galaxies' physical properties, including gas-phase metallicity, ionization properties, and electron temperature, and to correct the observed emission-line fluxes for dust attenuation.

We select a sample here requiring coverage of at least three rest-UV diagnostic lines among the UV \ciii, \civ, \oiii, and \heii\, detected with S/N $\ge 3$. For non-detected emission lines, we adopted the $1 \sigma$ flux uncertainty as its upper limit. These criteria yield a primary sample of eight galaxies at $z \sim 7$--$9$ (see Table \ref{tab:linefluxes} for spectroscopic redshifts): C3PO-21722, C3PO-45290 \footnote[1]{45290's UV spectrum is consistent with emission powered by stars, but, see \citet{Papovich2026}} (an LRD studied in \citet{Papovich2026}), C3PO-13264, C3PO-44218, C3PO-84478 (a Lyman-Alpha Emitter (LAE) studied in \citet{Hu2026}), C3PO-84579, and C3PO-97363 from pointing 2. We also included C3PO-101957 from pointing 1 in our \ciii\, rest-frame Equivalent Width (EW$_0$) versus metallicity plot because it has robust \ciii\, EW$_0$ and rest-optical emission-line measurements. For each object, we use the calibrated 2D spectrum (Figure \ref{fig: data_2d}) to determine emission-line detection. We present the extracted 1D spectral windows of the UV and optical emission lines that are used in this paper for diagnostics and physical properties in Figure \ref{fig: data_1d}.

We supplement this primary sample with literature measurements of rest-UV emission lines in high-redshift galaxies from \textit{JWST} and their local analogs from ground-based surveys for comparison in Section \ref{sec:results}.

\clearpage
\section{Characteristics of the High-z Sample}
\label{sec: characteristic}
\subsection{Line Measurement}
\label{sec: line_measurement}
\begin{deluxetable*}{lcccccccc}
\centerwidetable
\tablewidth{0pt}
\tabletypesize{\small}
\tablecaption{Measured emission-line fluxes and rest-frame equivalent widths for the C3PO sample.\label{tab:linefluxes}}
\tablehead{
\colhead{Parameter} & \colhead{101957} & \colhead{13264} & \colhead{97363} & \colhead{44218} & \colhead{21722} & \colhead{45290} & \colhead{84579} & \colhead{84478}
}
\startdata
Pointing & 1 & 2 & 2 & 2 & 2 & 2 & 2 & 2 \\
$z_{\rm spec}$ & 7.9232 & 7.8194 & 7.0301 & 7.4679 & 7.1670 & 8.3577 & 8.8791 & 8.9982 \\
\hline
\civ $\lambda1548$ & $<0.54$ & $1.18\pm0.60$ & $2.68\pm0.66$ & $3.65\pm2.32$ & $3.04\pm0.55$ & $2.27\pm0.58$ & $1.21\pm0.44$ & $2.13\pm0.39$ \\
\civ $\lambda1551$ & $<0.54$ & $7.05\pm0.60$ & $2.33\pm0.66$ & $6.17\pm2.33$ & $2.46\pm0.55$ & $2.72\pm0.58$ & $1.46\pm0.44$ & $1.63\pm0.39$ \\
\heii\ $\lambda1640$ & $<1.31$ & $5.02\pm0.78$ & $0.94\pm0.57$ & $3.50\pm1.70$ & $0.95\pm0.64$ & $0.88\pm0.69$ & $1.88\pm0.54$ & $<0.48$ \\
O\,{\sc iii}] $\lambda1666$ & $1.02\pm0.50$ & $4.10\pm0.80$ & $0.84\pm0.58$ & $3.66\pm1.82$ & $3.32\pm0.53$ & $4.44\pm0.64$ & $<0.63$ & $2.16\pm0.41$ \\
$[\ciii$ $\lambda1907$ & $1.13\pm0.60$ & $5.02\pm0.57$ & $1.57\pm0.57$ & $6.35\pm2.18$ & $3.30\pm0.46$ & $3.05\pm0.66$ & $1.52\pm0.52$ & $3.06\pm0.51$ \\
\ciii\ $\lambda1909$ & $1.01\pm0.61$ & $5.48\pm0.59$ & $1.64\pm0.60$ & $3.92\pm2.25$ & $2.24\pm0.48$ & $4.24\pm0.67$ & $0.80\pm0.52$ & $2.50\pm0.51$ \\
\hline
$EW_0(\ciii )$ [\AA] & $6.51\pm2.60$ & $11.11\pm0.87$ & \nodata\tablenotemark{\textsection} & $28.72\pm8.77$ & $10.96\pm1.31$ & $18.06\pm2.34$ & \nodata\tablenotemark{\textsection} & \nodata\tablenotemark{\textsection} \\
$EW_0(\oiii])$ [\AA] & $3.04\pm1.49$ & $2.95\pm0.58$ & $<2.97$ & $13.03\pm6.46$ & $5.02\pm0.80$ & $12.19\pm1.75$ & $<6.10$ & $19.65\pm3.75$ \\
\hline
\oii\ $\lambda3727$ & $1.60\pm0.31$ & $1.59\pm0.41$ & $1.27\pm0.53$ & $1.37\pm1.52$ & $0.89\pm0.48$ & $0.43\pm0.37$ & $0.34\pm0.25$ & $<0.20$ \\
\oii\ $\lambda3730$ & $1.48\pm0.39$ & $3.02\pm0.52$ & $<0.67$ & $4.33\pm1.94$ & $1.03\pm0.61$ & $<0.46$ & $<0.31$ & $<0.25$ \\
H$\gamma$ & $2.56\pm0.18$ & $7.21\pm0.25$ & $0.91\pm0.20$ & $2.67\pm0.73$ & $3.63\pm0.21$ & $3.33\pm0.30$ & $1.13\pm0.21$ & $2.81\pm0.20$ \\
$[\oiii]\lambda4364$ & $1.09\pm0.18$ & $2.78\pm0.25$ & $1.02\pm0.20$ & $<0.73$ & $1.05\pm0.21$ & $2.31\pm0.30$ & $0.11\pm0.21$ & $1.31\pm0.20$ \\
H$\beta$ & $5.79\pm0.23$ & $14.85\pm1.48$ & $2.86\pm0.22$ & $6.89\pm0.78$ & $8.25\pm0.24$ & $7.80\pm0.39$ & $2.67\pm0.29$ & $4.88\pm0.32$ \\
$[\oiii]\lambda4960$ & $10.70\pm {0.08}$ & $35.61\pm {0.50}$ & $6.23\pm {0.08}$ & $11.62\pm {0.26}$ & $14.99\pm {0.08}$ & $10.52\pm {0.13}$ & $4.09\pm {0.10}$\tablenotemark{*} & $ {11.83\pm0.11}$\tablenotemark{*} \\
$[\oiii]\lambda5008$ & $31.89\pm0.23$ & $106.11\pm1.48$ & $18.56\pm0.22$ & $34.61\pm0.78$ & $44.68\pm0.24$ & $31.35\pm0.39$ & $12.18\pm0.29$\tablenotemark{*} & $ {35.26\pm0.32}$\tablenotemark{*} \\
\enddata
\tablecomments{Fluxes are observed values, before internal dust correction, in units of $10^{-19}\,\mathrm{erg\,s^{-1}\,cm^{-2}}$. Quoted errors are formal $1\sigma$ fit uncertainties. Entries with flux-to-error ratios below three should not be interpreted as secure detections.}
\tablenotetext{*}{Measurements use the extended-wavelength spectra.}
\tablenotetext{\text{\textsection}}{C III] equivalent widths are omitted when the local continuum has $\mathrm{S/N}<1$ or is otherwise poorly constrained, preventing a reliable EW measurement. The fitted local continuum of 84579 is negative.}
\end{deluxetable*}

We measure the rest-frame UV and optical emission-line fluxes and EWs from the one-dimensional spectra presented above in Figure \ref{fig: data_1d}. We model the emission lines with Gaussian profiles using an unweighted least-squares procedure. We assume the lines are unresolved, and fix the Gaussian widths to the theoretical resolution curve. We fit the components of each doublet simultaneously and jointly fit  neighboring emission features when their central wavelengths are separated by less than 50 \AA, such as \hb\, and [\oiii]$\lambda\lambda 4960, 5008$. Within each doublet or group, the component centroids are tied to a fitted redshift. The \heii\, centroid is fixed at the systematic spectroscopic redshift derived from fitting \oiii\,$\lambda\lambda4960, 5008$ and \hb\, lines simultaneously. We approximate the continuum with a linear function within the local window of $\sim 100\,\text{\AA}$, excluding the emission feature.

We compute the $\mathrm{EW}_0$ as the ratio of the integrated line flux to the local continuum flux density, divided by $(1+z)$. We obtain the continuum from the best-fit linear continuum model at the line center, or at the midpoint for the doublet. We calculate the EW uncertainties by propagating the line-flux uncertainty while treating the fitted continuum as fixed. The resulting line and EW measurements are summarized in Table~\ref{tab:linefluxes}.

We then correct the measured line fluxes for internal dust attenuation using the \citet{Cardelli1989} extinction curve with $R_V = 3.1$, adopting the color excess derived from the Balmer decrement. We use the \texttt{getEmissivity} routine of the \textsc{PyNeb} H{\sc i} recombination atom to calculate the intrinsic ratio, $(\hg/\hb)_{\rm int}$. We assume Case B recombination at an electron temperature $T_e \simeq 18{,}000\,\mathrm{K}$ and an electron density $n_H \simeq 20,000\,\mathrm{cm}^{-3}$.\footnote[2]{These values are adopted only to calculate the intrinsic Balmer ratio, which depends weakly on temperature and density under Case B recombination \citep{Osterbrock1989}.} We compare this intrinsic ratio with the observed $(\hg/\hb)_{\rm obs}$ ratio to derive $E(B-V)$. We apply the resulting correction to the measured emission-line fluxes from both gratings before deriving the physical properties described in the next section.

\subsection{Physical Properties}

We get the nebular physical conditions and abundance ratios using Monte Carlo propagation of the measured emission-line flux uncertainties. For each galaxy, we sample the posterior distributions of electron density, electron temperature, oxygen abundance, and carbon abundance using the observed density- and temperature-sensitive line ratios as constraints.  For galaxies with insufficient emission-line coverage to constrain $T_e$ and/or $n_H$ directly, we adopt fiducial values of $T_e = 18{,}000$\,K and/or $n_H = 20,000\,\mathrm{cm^{-3}}$. 

\subsubsection{Electron Density}
\label{sec:ne}
We adopt \ciii-based electron densities $n_e(\ciii)$, for our physical-property measurements because this diagnostic probes the higher-ionization gas relevant to our rest-UV analysis. Producing $\mathrm{C^{2+}}$ requires photons with energies above 24.4\ eV, compared with 13.6 and 10.4\,eV for producing $\mathrm{O^{+}}$ and $\mathrm{S^{+}}$, respectively. We derive the electron density in this doubly ionized carbon zone from the dust-corrected flux ratio of the two \ciii\, doublet components,
\begin{equation}
    R_{\rm C\,III]} = \frac{F(\mathrm{[C\,III]}\lambda1907)}{F(\mathrm{C\,III]}\lambda1909)},
\end{equation}
using the \texttt{getTemDen} function in the \textsc{PyNeb} package \citep{Luridiana2015}. We calculate the initial density at $T_e = 18{,}000$ K, estimate $T_e$ from the rest-frame optical [\oiii] line ratios at this density (Section~\ref{sec:te}), and then update the density using the resulting temperature. For Monte Carlo processes in which the temperature diagnostic does not return a valid solution due to a weak auroral line or a broad auroral line component, we assume a fiducial $T_e = 18{,}000$\,K.

The values of $n_e(\ciii)$ in our samples span a wide range (as shown in Table \ref{tab:physical_properties}). For the six galaxies from our sample with reported densities, the average values span approximately $1.8\times10^3$--$4.9\times10^4\,\mathrm{cm^{-3}}$. For comparison, \citet{Topping2025} find a median of $\sim 1.4 \times 10^4$ for their $z>5$ samples with \ciii\, detections. \citet{Maseda2017} report a median $\log[n_e/(\mathrm{cm^{-3}})] = 3.2$ for their intermediate-redshift \ciii\ emitters, assuming $T_e = 10{,}000$ K. Local analogs from CLASSY have \ciii-based densities of $\sim 10^4 \mathrm{cm^{-3}}$ \citep{Mingozzi2024}. Our measured values are comparable with both local high-redshift analogues and other intermediate-to-high-redshift \ciii\, emitters. Differences in sample selection and diagnostic assumptions prevent this comparison alone from establishing density evolution. However, we show that in these higher-ionized galaxies, the electron density is much higher than what models commonly assume.

We caution that the density $n_e(\ciii)$ is an effective electron density inferred from the integrated emission of the doubly ionized carbon zone, whereas the model input $n_H$ specifies the hydrogen number density. Although these quantities are closely related in ionized gas, their comparison depends on the ionization state and density structure. Low-ionization diagnostics also show elevated densities at high redshift. For the \sii\ diagnostic, local SDSS star-forming galaxies have a characteristic electron density of $n_e(\sii)\sim26\,\mathrm{cm^{-3}}$ \citep{Sanders2016}, compared with the median $n_e(\sii)\sim480\,\mathrm{cm^{-3}}$ reported at $z>4$ by \citet{Topping2025}. For the \oii\ diagnostic, characteristic densities increase from $n_e(\oii)=114^{+28}_{-27}\,\mathrm{cm^{-3}}$ at $z\sim1.5$ \citep{Kaasinen2017} to $n_e(\oii)\sim225\,\mathrm{cm^{-3}}$ at $z\sim2.3$ \citep{Sanders2016} and $\gtrsim300\,\mathrm{cm^{-3}}$ at $z=4$--9 \citep{Isobe2023}. These measurements support an increase in typical low-ionization-zone densities with redshift, although the samples differ in selection and exhibit substantial scatter. The \ciii\ diagnostic can probe denser gas in a higher-ionization zone, and the densities inferred for our sample motivate exploring high-density photoionization grids.

\subsubsection{Electron Temperature}
\label{sec:te}
We derive the electron temperature using the direct $T_e$ method, applied to the rest-frame optical [\oiii]\ emission lines. We use the temperature-sensitive ratio between the auroral [\oiii]\,$\lambda$4364 line and the strong nebular [\oiii]\,$\lambda\lambda$4960,5008 lines,
\begin{equation}
R_{\rm [O\,III]} =
\frac{F(\mathrm{[O\,III]}\lambda4364)}
{F(\mathrm{[O\,III]}\lambda4960) + F(\mathrm{[O\,III]}\lambda5008)} .
\end{equation}
All fluxes are corrected for dust attenuation before calculating this ratio. We calculate $T_e$ from $R_{\oiii}$ using the \texttt{getTemDen} function in \textsc{PyNeb}, assuming the electron density derived from \ciii\, (Section~\ref{sec:ne}). We subsequently update the \ciii-based density using the resulting temperature. The resulting $T_e$ values (listed in Table~\ref{tab:physical_properties}) characterize the high-ionization nebular gas in the O$^{2+}$ zone, and we adopt them when deriving the oxygen abundance associated with this ionization state (Section~\ref{sec:oh}).

\subsection{Oxygen Abundance Measurement}
\label{sec:oh}
We derive the gas-phase oxygen abundance using 
\begin{equation}
\frac{\mathrm{O}}{\mathrm{H}} =
\frac{\mathrm{O}^{+}}{\mathrm{H}^{+}} + \frac{\mathrm{O}^{2+}}{\mathrm{H}^{+}},
\end{equation}
assuming that neutral oxygen and $\mathrm{O^{3+}}$ or higher ionization states contribute negligibly to the oxygen abundance within the ionized gas (as demonstrated in Figure~\ref{fig:ion_structure}). 
The doubly ionized abundance, $\mathrm{O}^{2+}/\mathrm{H}^{+}$, is calculated from the strong [\oiii]\,$\lambda\lambda$4960,5008 nebular lines relative to H$\beta$, using $T_e$ from Section~\ref{sec:te}. The singly ionized abundance, $\mathrm{O}^{+}/\mathrm{H}^{+}$, is calculated from the \oii\,$\lambda\lambda$3727,3730 doublet relative to H$\beta$, adopting the temperature relation from \citet{Garnett1992},
\begin{equation*}
    T_e(\hbox{O {\sc ii}}) = 0.7\ T_e(\oiii) +3{,}000K
\end{equation*}
as an approximation for the low-ionization-zone. We calculate both ionic abundances using the \texttt{getIonAbundance} function in \textsc{PyNeb}, with dust-corrected fluxes and the updated \ciii-based electron density. 

\subsubsection{C/O Abundance}
Carbon-to-oxygen abundance ratios provide insight into chemical enrichment history and also test the abundance patterns assumed in the rest-UV photoionization models. Following the temperature-sensitive formalism described above, we estimate the ionic abundance ratio $\mathrm{C}^{2+}/\mathrm{O}^{2+}$ using the relation from \citet{Erb2010},
\begin{equation}
    \frac{\mathrm{C}^{2+}}{\mathrm{O}^{2+}} =
0.15\,\exp(-1.1054/t)\,
\frac{I(\ciii\ \lambda\lambda1907,1909)}
     {I(\oiii]\ \lambda\lambda1661,1666)},
\end{equation}
where $t = T_e(\oiii)/10^4$\,K, and $I$ denotes the dust-corrected line flux. Because \oiii]\,$\lambda$1661 is intrinsically weak and typically has low S/N, we estimate the flux of this line, when undetected, by adopting the theoretical intrinsic ratio \oiii]\,$\lambda1661/ \oiii]\,\lambda1666 \approx 1/2.5$ from \citet{Fischer1985}.

Even though \civ\, is detected for most of our targets, it cannot be taken as a clean nebular line according to \citet{Berg2019b} because \civ\, emission can result from a combination of nebular and stellar features. We therfore use an ionization correction rather than directly adding a $\mathrm C^{3+}$ abundance from \civ. The total elemental ratio is recovered using:
\begin{equation}
    \frac{\mathrm{C}}{\mathrm{O}} =
    \frac{\mathrm{C}^{2+}}{\mathrm{O}^{2+}} \times \mathrm{ICF},
    \qquad
    \mathrm{ICF} \equiv
    \left[\frac{X(\mathrm{C}^{2+})}{X(\mathrm{O}^{2+})}\right]^{-1}
    = \frac{X(\mathrm{O}^{2+})}{X(\mathrm{C}^{2+})},
\end{equation}
using the ionization correction factor (ICF), where $X(\mathrm{C}^{2+})$ and $X(\mathrm{O}^{2+})$ are the ionization fractions of $\mathrm{C}^{2+}$ and $\mathrm{O}^{2+}$ relative to their total elemental abundances. In high-ionization nebulae where our galaxies live, oxygen is almost entirely doubly ionized, but carbon is distributed between $\mathrm{C}^{2+}$ and $\mathrm{C}^{3+}$; the ICF therefore corrects
primarily for carbon residing in $\mathrm{C}^{3+}$.

To determine the carbon ICF, we adopt the approach of \citet{Berg2019b}, based on a grid of Cloudy photoionization models powered by \textsc{BPASS} binary stellar population \citep{Eldridge2017}. The ionization parameter $U$ measures the density of hydrogen-ionizing photons relative to the hydrogen density and is defined as
\begin{equation}
\label{eqn:U}
    U = \frac{Q_{\rm H}}{4\pi r^2 cn_{\rm H}}
\end{equation}
where $Q_H$ is the number of ionizing photons emitted per second, $c$ is the speed of light $2.9979 \times 10^{10} \rm cm~s^{-1}$, $r$ is the inner radius, and $n_{\rm H}$ is the number density of hydrogen. We first estimate $\log U$ from the dust-corrected optical excitation ratio $\mathrm{O_{32}} \equiv I([\mathrm{O\,III}]\,\lambda5008)/I([\mathrm{O\,II}]\,\lambda\lambda3727,3730)$ using the metallicity-dependent calibration given in Table~3 of \citet{Berg2019b}. We linearly interpolate the tabulated coefficients using the measured $12+\log(\mathrm{O/H})$ of each galaxy. We then use the $\log U$ derived from $O_{32}$ to compute the carbon ICF with the corresponding metallicity-dependent calibration from the same model grid, again interpolating the coefficients in metallicity and restricting $\log U$ to the calibrated range $[-3,-1]$. We combine the resulting ICF with the measured $\mathrm{C}^{2+}/\mathrm{O}^{2+}$ ionic abundance ratio to obtain the total C/O abundance.

Both the $O_{32}$--$U$ calibration and the carbon ICF assume stellar photoionization. An AGN contribution could alter the carbon and oxygen ionization fractions and the $O_{32}$--$U$ relation, introducing an additional systematic uncertainty in the inferred C/O abundances.

\begin{figure*}[!ht]
    \centering
    \includegraphics[width=1.0\textwidth]{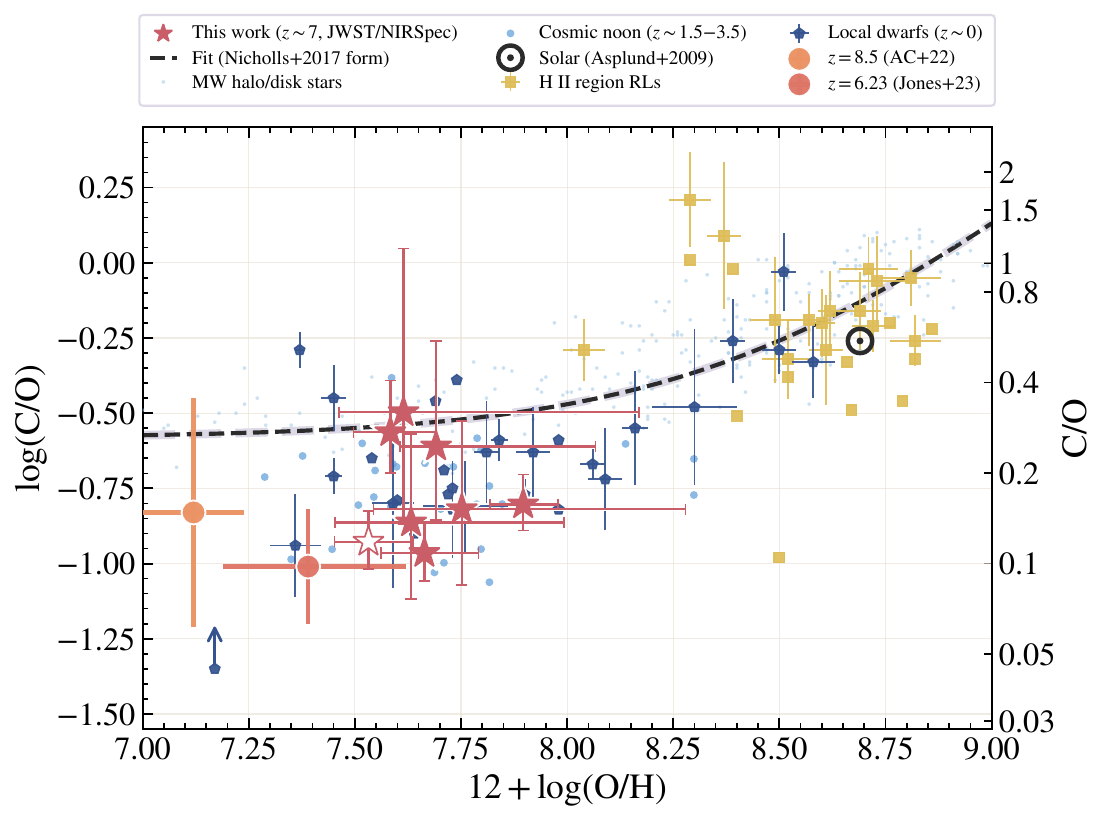}
    \caption{The relation between C/O and oxygen abundance, 12 + $\log\mathrm{O/H}$.  The C3PO $z\sim7-9$ galaxies are shown as filled, red stars (the open red star indicates the LRD, C3PO-45290).  Other points show data from the literature, including  Milky Way halo and disk stars (small light-blue dots; \citealt{Gustafsson1999, Akerman2004, Fabbian2009, Nissen2014}), galaxies at $z\sim1.5$--$3.5$ (larger blue circles; \citealt{Fosbury2003, Erb2010, Christensen2012, Bayliss2014, James2014, Stark2014, Steidel2016, Vanzella2016, Amorin2017, Berg2018, Mainali2020, Matthee2021, Rigby2021, Iani2023}), local dwarf galaxies at $z\sim0$ (dark-blue pentagons; \citealt{Berg2016, Berg2019b, Garnett1995, Garnett1999}), and \hii-region recombination-line measurements (gold squares; \citealt{Tsamis2003, GarciaRojas2004, GarciaRojas2005, GarciaRojas2006, Esteban2009}). We also include two \textit{JWST} EoR galaxies from \citet{ArellanoCordova2022} at $z=8.5$ (orange circle) and \citet{Jones2023} at $z=6.23$ (salmon circle). The solar C/O and $12+\log(\mathrm{O/H})$ values from \citet{Asplund2009} are marked by the black $\odot$ symbol, with dotted lines indicating the solar $\log(\mathrm{C/O})=-0.26$ and $12+\log(\mathrm{O/H}) = 8.69$. The dashed black line shows the C/O--$12+\log(\mathrm{O/H})$ relation of \citet{Nicholls2017}.}
    \label{fig:co_oh_relation}
\end{figure*}

Figure~\ref{fig:co_oh_relation} compares our measurements with literature abundance measurements spanning Milky Way halo and disk stars, local ($z\sim0$) dwarf galaxies, recombination-line measurements of $z\sim0$ \hii\ regions, cosmic-noon galaxies with $z\approx 2-3$, and recent \textit{JWST} measurements of galaxies in the EoR (see figure caption for references). Our galaxies have C/O that spans approximately $0.11$--$0.32$, well below the solar reference $\mathrm{C/O} = 0.55$ shown by the dotted line \citep{Asplund2009}. Our sample and other EoR galaxies occupy this low-C/O, low-O/H regime, motivating the adoption of lower C/O abundances when constructing photoionization simulations for reionization-era galaxies.

The measurement results for these physical properties are listed in Table~\ref{tab:physical_properties}.

\begin{deluxetable*}{lcccccccc}
\centerwidetable
\tablewidth{0pt}
\tabletypesize{\small}
\tablecaption{Derived nebular physical properties for the C3PO sample.\label{tab:physical_properties}}
\tablehead{
\colhead{Parameter} & \colhead{101957} & \colhead{13264} & \colhead{97363} & \colhead{44218} & \colhead{21722} & \colhead{45290} & \colhead{84579} & \colhead{84478}
}
\startdata
$n_e(\mathrm{CIII]})$ [$10^4$ cm$^{-3}$] & ${2.0}^{+{6.3}}_{-{0.3}}$ & $2.8^{+1.1}_{-1.0}$ & $2.6^{+4.8}_{-{1.3}}$ & \nodata & ${2.0}^{+{0.0}}_{-{1.6}}$ & $4.9^{+2.9}_{-2.0}$ & \nodata & ${1.9}^{+{0.9}}_{-{1.3}}$ \\
$T_e$ [$10^4$ K] & $2.0^{+{0.2}}_{-0.3}$ & $1.6^{+0.1}_{-0.1}$ & \nodata & $1.8^{+0.1}_{-0.7}$ & $1.7^{+0.2}_{-0.2}$ & \nodata & $1.7^{+0.1}_{-{0.6}}$ & ${2.0}^{+0.2}_{-0.2}$ \\
$12+\log(\mathrm{O/H})$ & ${7.69}^{+{0.31}}_{-{0.14}}$ & $7.90^{+0.08}_{-0.08}$ & ${7.71}^{+{0.36}}_{-{0.10}}$ & ${7.89}^{+{0.49}}_{-{0.27}}$ & ${7.70}^{+0.13}_{-0.10}$ & $7.53^{+0.11}_{-0.08}$ & ${7.66}^{+0.55}_{-{0.18}}$ & ${7.59}^{+{0.10}}_{-0.09}$ \\
$\mathrm{C^{2+}/O^{2+}}$ & ${0.13}^{+0.13}_{-0.06}$ & $0.14^{+0.03}_{-0.02}$ & $0.27^{+0.33}_{-0.11}$ & ${0.16}^{+0.16}_{-0.07}$ & $0.10^{+0.02}_{-0.02}$ & $0.10^{+0.02}_{-0.02}$ & $0.30^{+0.75}_{-0.18}$ & $0.16^{+0.04}_{-0.03}$ \\
ICF & $1.01^{+0.02}_{-0.02}$ & $1.15^{+0.03}_{-0.02}$ & $0.93^{+0.05}_{-0.04}$ & $0.91^{+0.06}_{-0.05}$ & $1.10^{+0.08}_{-0.04}$ & $1.14^{+0.12}_{-0.06}$ & $1.07^{+0.09}_{-0.05}$ & $1.66^{+0.59}_{-0.25}$ \\
$\log U$\tablenotemark{*} & ${-2.21}^{+{0.15}}_{-{0.08}}$ & $-1.68^{+0.09}_{-0.08}$ & ${-2.39}^{+{0.16}}_{-{0.12}}$ & ${-2.46}^{+{0.20}}_{-{0.14}}$ & ${-1.90}^{+{0.20}}_{-0.14}$ & $-1.86^{+0.22}_{-0.15}$ & ${-1.92}^{+{0.34}}_{-{0.22}}$ & ${-1.18}^{+0.36}_{-0.24}$ \\
$\mathrm{C/O}$ & $0.14^{+0.13}_{-0.06}$ & $0.16^{+0.04}_{-0.03}$ & ${0.24}^{+0.31}_{-0.11}$ & $0.15^{+0.15}_{-0.07}$ & $0.11^{+0.03}_{-0.02}$ & $0.12^{+0.03}_{-0.02}$ & $0.32^{+{0.79}}_{-0.18}$ & $0.27^{+0.13}_{-0.07}$ \\
\enddata
\tablecomments{Values are medians, with uncertainties spanning the 16th--84th percentiles of the finite Monte Carlo realizations for each quantity. When a diagnostic fails in a Monte Carlo realization, we adopt $T_e=18{,}000\,\mathrm{K}$ or $n_e=2\times10^4\,\mathrm{cm^{-3}}$, as appropriate. These values are the rounded medians of the six reported temperatures ($17{,}542\,\mathrm{K}$) and six reported densities ($20{,}241\,\mathrm{cm^{-3}}$), providing fiducials representative of our sample. The uncertainties are conditional on these fixed values and do not include uncertainty in the fiducials. Density medians coincident with the fiducial are not independent density measurements. The listed temperature is $T_e(\mathrm{O\,III})$; we adopt $T_e(\mathrm{O\,II})=0.7T_e(\mathrm{O\,III})+3000\,\mathrm{K}$.}
\tablenotetext{*}{Derived following \citet{Berg2019b}, using the O32$=I([\mathrm{O\,III}]\lambda5008)/I([\mathrm{O\,II}]\lambda\lambda3727,3730)$ ratio.}
\end{deluxetable*}

\section{Photoionization Models}
\label{sec:models}
In this section, we describe star-forming and AGN photoionization models and their dependence on C/O and gas density. We compare these models with the observations in Section~\ref{sec:results}.

\subsection{Star-Forming Models}
\label{sec: SF}
We compute nebular emission models using the photoionization code \textsc{Cloudy} C25.00 \citep{Gunasekera2025-CLOUDY}, with ionizing spectra from the Binary Population and Spectral Synthesis models (\textsc{BPASS} v2.2.1; \citealt{Eldridge2017, Stanway2018}). We use the \textsc{BPASS} models that include the effects of binary star evolution, in which mass transfer, stellar stripping, and mergers can extend the production of hard ionizing photons to later stellar ages compared to single-star populations. These effects are important for young, actively star-forming galaxies, where the hardness of the ionizing spectrum directly shapes the strength of rest-frame UV nebular lines \citep{Eldridge2009, Eldridge2022, Steidel2016, Feltre2016}. For comparison, we also include stellar spectra based on the \citet{BruzualCharlot2003} (BC03/MILES) models, since similar single-star stellar population synthesis models have been widely adopted in previous diagnostic grids and analyses of observational data.

For each stellar model, we generate a single-burst stellar population spectrum using \textsc{Bagpipes} \citep{Carnall2018}. This spectrum is then used as the incident radiation field for \textsc{Cloudy}. We model the nebular gas as a spherical \hii\ region surrounding the young stellar population, with the inner radius fixed to $r=10^{19}\,\mathrm{cm}$. The ionization parameter, $U$, is varied by changing the normalization of the ionizing radiation field at fixed geometry and gas density.

Our model grids are designed to isolate the effects of C/O abundance ratio and nebular density on rest-frame UV diagnostics. We consider ${\rm C/O}=0.12$, 0.18, and 0.36, and $n_{\rm H}=100$, 1000, and $10000\,\mathrm{cm^{-3}}$. The parameters of the \textsc{BPASS}-based grids are summarized in Table~\ref{tab:sf_grid}. Here, $n_{\rm H}$ is the input hydrogen number density used by \textsc{Cloudy}, whereas $n_e$ is the electron density calculated within the nebula. In practice, $n_e$ is approximately equal to $n_{\rm H}$ in the ionized gas, with a contribution from ionized helium. We use the \textsc{Bagpipes} default values, ${\rm C/O}=0.36$ and $n_{\rm H}=100\,\mathrm{cm^{-3}}$, as a fiducial reference case.

\begin{table}
\centering
\begin{tabular}{ll}
\hline
Parameter & Adopted values \\
\hline
Stellar population & \textsc{BPASS} v2.2.1, binary \\
IMF & \texttt{imf135\_300} \\
Star formation history & Instantaneous burst \\
$\log(t/\mathrm{yr})$ & 6.0--7.4, in steps of 0.1 \\
C/O & 0.12, 0.18, 0.36 \\
$\log(n_{\rm H}/\mathrm{cm^{-3}})$ & 2.0, 3.0, 4.0 \\
$\log U$ & $-4.0$ to $0.0$, in steps of 0.5 \\
Metallicity, $Z/Z_\odot$ & 0.0005, 0.005, 0.05, 0.10, 0.15, 0.20, \\
& 0.30, 0.40, 0.50, 0.70, 1.00, 1.50, 2.00 \\
\hline
\end{tabular}
\caption{Parameter space of the \textsc{BPASS}-based star-forming photoionization grids. C/O is a number abundance ratio, and $n_{\rm H}$ is the hydrogen number density. Stellar and nebular metallicities are tied, with $Z_\odot=0.02$. For the UV diagnostic comparisons, we adopt a stellar age of 1 Myr and evaluate $Z/Z_\odot=0.001$, 0.003, 0.01, 0.05, and 0.20 at $\log U=-3$, $-2$, $-1$, and 0, interpolating within the stored metallicity grid where necessary.}
\label{tab:sf_grid}
\end{table}

We express the model metallicities in solar units, adopting $Z_\odot=0.02$ as the normalization for the stellar-population grids. For comparison with the observed oxygen abundances, we map the model metallicities onto an adopted solar oxygen abundance scale using $12+\log(\mathrm{O/H})=8.69+\log(Z/Z_\odot)$ \citep{Asplund2009}. This plotting convention differs slightly from the oxygen normalization of 8.71 in the nebular input abundance prescription. Our \textsc{BPASS} grid spans $\log U=-4.0$ to 0.0 and metallicities from $5\times10^{-4}\,Z_\odot$ to $2\,Z_\odot$, including all 13 metallicities sampled by the adopted stellar models. For the UV diagnostic comparisons, we use a 1 Myr burst and the parameter subset specified in Table~\ref{tab:sf_grid}.

For the elemental abundances, C/H is varied explicitly with C/O, such that $\mathrm{C/H=(C/O)\times(O/H)}$, while nearly all other heavy elements are scaled linearly with oxygen abundance. The exceptions are helium and nitrogen, whose abundances are metallicity-dependent \citep[see][]{Carnall2018}. This setup allows us to test whether the displacement of high-redshift galaxies in UV diagnostic diagrams can be explained by physically motivated changes in the abundance pattern and nebular density, rather than requiring an additional AGN ionizing component.

\subsubsection{Ionization Structure of Carbon and Oxygen}
\label{sec:ionization_structure}
\begin{figure*}[!ht]
    \centering
    \includegraphics[width=\textwidth]{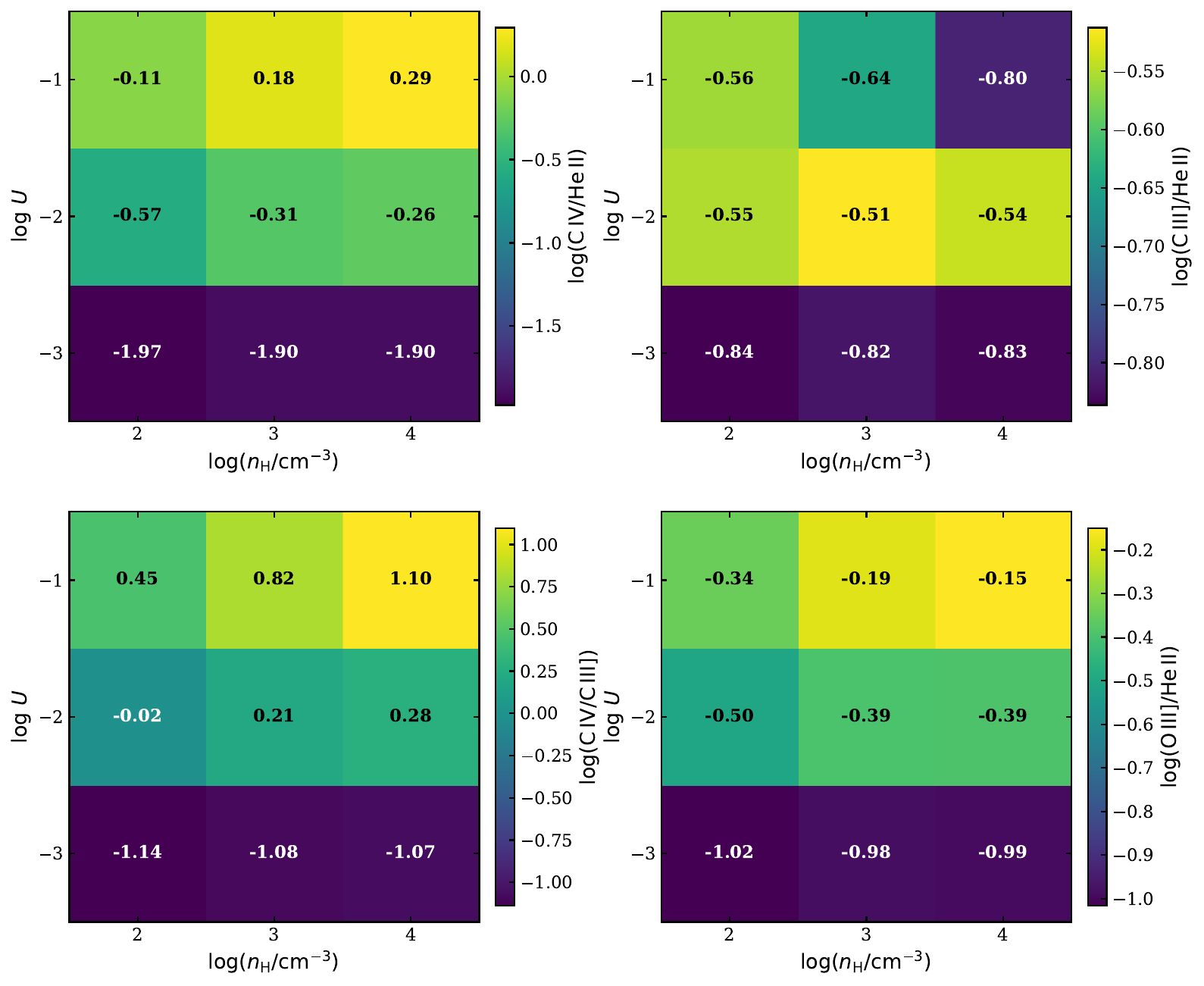}
    \caption{Line-ratio grids for photoionization models spanning $\log U=-1,-2,-3$ and $\log(n_{\rm H}/{\rm cm^{-3}})=2,3,4$. Each heatmap shows the predicted $\log$ ratio for one UV diagnostic: \civ/\heii\ (top left), \ciii/\heii\ (top right), \civ/\ciii\ (bottom left), and \oiii]/\heii\ (bottom right). The grids illustrate how each line ratio responds to changes in density and ionization parameter.  As noted in the text, $\log Q \sim \log U +\log n_H$, so the diagonal from upper left to lower right traces a fixed ionizing production rate, $Q$.  At fixed $\log U$, increasing $\log n_H$ requires a similar increase in $\log Q$.  }
    \label{fig:line_ratio_heatmap}
\end{figure*}
\begin{figure*}[!ht]
    \centering
    \includegraphics[width=\textwidth]{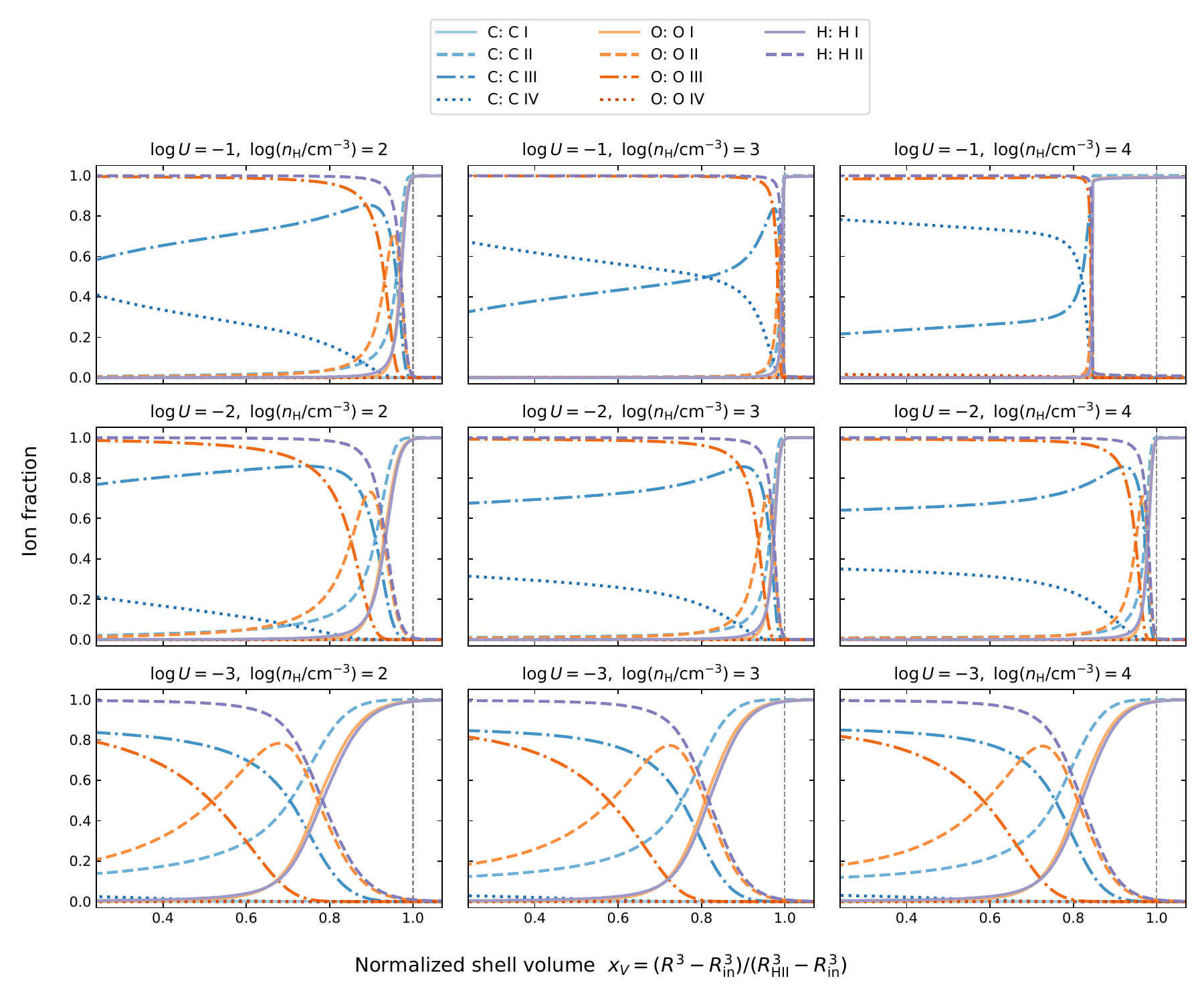}
    \caption{Radial ionization structure of the spherical photoionization models. The lines show the ionization fractions for different transitions of carbon (blue lines), oxygen (orange lines), and hydrogen (purple lines) as a function of the normalized shell volume, $x_V=(R^3-R_{\rm in}^3)/(R_{\rm HII}^3-R_{\rm in}^3)$. Here $R=R_{\rm in}+d$, where $d$ is the depth from the illuminated face and $R_{\rm in}=10^{19}\,\mathrm{cm}=3.24\,\mathrm{pc}$. Each panel shows models spanning different combinations of $\log U=-1$, $-2$, $-3$ (rows) and $\log(n_{\rm H}/\mathrm{cm}^{-3})=2$, 3, 4 (columns). Dashed grey vertical lines mark the adopted hydrogen ionization boundary at $R_{\rm HII}$ ($x_V=1$), where the \hii\ fraction first falls to 1 per cent. Equal intervals in $x_V$ correspond to equal volume increments within the spherical shell. For each element, solid lines show the neutral species, dashed lines the singly ionized state, dash-dotted lines the doubly ionized state, and dotted lines the triply ionized state.}
    \label{fig:ion_structure}
\end{figure*}

To isolate the dependence of line ratios on the ionizing field and the gas density, we examine how the predicted ratios vary with different combinations of $\log U$ and $n_{\mathrm{H}}\approx n_e$ values in Figure~\ref{fig:line_ratio_heatmap}.  Figure~\ref{fig:ion_structure} shows the nebular structure for each case).  Each panel shows a unique diagnostic of \civ/\heii, \ciii/\heii, \civ/\ciii, and \oiii]/\heii, with color encoding the logarithm of the line flux ratio. All ratios quoted in this section are outputs of the photoionization grid described above, with fixed C/O = 0.12. Following Equation~\ref{eqn:U}, the ionization scales as $U \propto Q_{\rm H}/(n_{\rm H}\,c)$ at fixed geometry and an inner radius of $R$ (fixed at $10^{19}$cm). Therefore, $\log U + \log(n_e)$ traces a constant ionizing-photon production rate $Q_{\rm H}$. The diagonal of each panel, running from top-left to lower-right, thus corresponds to fixed $Q_{\rm H}$, with only the relative values of $\log U$ and $n_{\rm H}$ changing. 

At fixed $n_H$, \civ/\heii, \civ/\ciii, and \oiii]/\heii\ all increase monotonically with $\log U$. A higher $\log U$ delivers more ionizing photons per hydrogen atom, shifting the ISM toward higher ionization states and thus extending the C$^{3+}$ and O$^{2+}$ zones deeper into the nebula at the expense of their lower-ionization counterparts. Because \heii\ is a recombination line that traces the hard ionizing photon field (photons with energy $\geq 54.4~ \rm eV$) while being relatively insensitive to metallicity and abundance ratios, the high-ionization metal lines rise relative to \heii\ as $\log U$ increases. We also observe that, along the constant-$Q_{\rm H}$ direction, the same ratios decline toward higher $n_{\rm e}$ and lower $\log U$: although the total ionizing budget is the same, the lower photon-to-gas ratio reduces the high-ionization fraction. The UV ratios are therefore set not only by the ionizing photon rate alone, but also by the photon density relative to the gas density.

Around $\log U\sim-2$, the predicted UV diagnostic line ratios are much less sensitive to gas density when increasing from $n_{\rm H}=10^3$ to $10^4~{\rm cm}^{-3}$ than from $n_{\rm H}=10^2$ to $10^3~{\rm cm}^{-3}$, as shown in Figure~\ref{fig:line_ratio_heatmap}. The corresponding ionization structures likewise remain more similar between $n_{\rm H}=10^3$ and $10^4~{\rm cm}^{-3}$ (Figure~\ref{fig:ion_structure}).

The gas density $n_H$ does play a significant role. At fixed $\log U$, input ionizing spectrum, geometry, and inner radius ($R = 10^{19}$\,cm), the incident radiation field is held fixed while $n_H$ varies. Increasing $n_H$ from $10^{2}$ to $10^{4}\,{\rm cm^{-3}}$ produces a density-dependent enhancement of the high-ionization carbon ratios, converting C$^{2+}$ to C$^{3+}$ more efficiently (see Figure~\ref{fig:ion_structure}). This is directly reflected in our line ratio grids. $\log(\civ/\ciii)$ rises from $0.45$ to $1.10$ and $\log(\civ/\heii)$ from $-0.11$ to $0.29$ at $\log U=-1$ across this density range, showing a large shift of carbon from C$^{2+}$ to C$^{3+}$. Because the ionizing spectrum, geometry, and inner radius remain unchanged, this trend does not represent a hardening of the radiation field but instead results from the density dependence of the recombination rates and the collisionally excited populations. 

The $\ciii/\heii$ ratio shows a more complicated behavior. The ratio reaches a maximum near $\log U \simeq -2$ and $\log(n_{\rm H}/\mathrm{cm^{-3}}) \simeq 3$, and declines toward both higher and lower $\log U$ and $\log(n_{\rm H}/\mathrm{cm^{-3}})$. This reflects the balance between the relative strength of the two lines: \heii\, is a recombination line set by hard ionization ($\geq 54.4~{\rm eV}$), whereas \ciii\, is a semi-forbidden, collisionally excited doublet whose emissivity scales as $n_{e}\,n({\rm C}^{2+})$ and is strongly temperature-sensitive. Two effects suppress $\ciii/\heii$ on either side of the peak. Toward higher $\log U$, carbon is converted into C$^{3+}$, depleting the C$^{2+}$ reservoir that powers \ciii\, emission while the \heii\, zone grows. Towards lower $\log U$, the model-predicted electron temperature falls from $T_{\rm e}\sim1.75\times10^{4}$~K at $\log U=-1$ to $\sim1.08\times10^{4}$~K at $\log U=-3$ for fixed $n_{\rm H}=1000~\mathrm{cm^{-3}}$. Since the \ciii\ excitation energy is $\approx6.5$~eV, this drop suppresses the collisionally excited line. The density dependence is comparatively weak. The $\ciii\,\lambda1907/\lambda1909$ doublet is a classic density diagnostic \citep{Jaskot2016}, but over the range probed here it is chiefly the internal  $\lambda1907/\lambda1909$ ratio that responds: the low-critical-density $\lambda1907$ component begins to be collisionally de-excited near $n_{\rm e}\sim10^{4}$--$10^{5}\,{\rm cm^{-3}}$ while $\lambda1909$ remains unaffected, so the total \ciii\, emissivity relative to the recombination-powered \heii. Together, these impacts place the maximum of $\ciii/\heii$ at intermediate $\log U \simeq -2$ and $n_{\rm H}\sim10^{3}\,{\rm cm^{-3}}$.

Figure \ref{fig:ion_structure} shows the ionization structure of hydrogen, carbon, and oxygen for the same grid of the Cloudy model (C/O = 0.12). The ion fraction is plotted as a function of normalized depth to the hydrogen ionization boundary (defined as x$_{\mathrm{HII}}=0.01$). The higher ionization species C$^{2+}$ and C$^{3+}$ occupy the inner, hard-ionization zone, while the low-ionization species are confined to progressively thinner shells toward the hydrogen ionization boundary. At higher $\log U$, more photons are available per particle, so the high-ionization species C$^{3+}$ and O$^{2+}$ reach deeper into the cloud before the photons are used up, expanding the higher-ionization species layer. The fraction for higher-ionization species is also higher across the nebula. As $n_H$ increases, the gas in the nebula is denser, squeezing the lower-ionization shells into a thinner region.

\subsubsection{UV Emission Line Diagnostic Grids}
UV emission-line diagnostic diagrams are widely used tools for separating and classifying galaxies as star-forming or AGN \citep{VillarMartin1997, Allen1998, Feltre2016, Gutkin2016, Hirschmann2019, Hirschmann2023, Plat2026}. Their predicted loci and classification boundaries depend on the adopted ionizing spectra, abundance patterns, and gas densities. 
To illustrate these dependencies, we compare the emission line ratios from our Cloudy star-forming models with the \citet{Gutkin2016} star-forming locus, the \citet{Plat2026} AGN model locus, and the \citet{Hirschmann2023} boundaries.   For these UV diagnostic plots, we show our results for a young stellar population with age of $1$ Myr.  We show how our model grids shift across the diagnostic planes as $n_H$ and C/O change. We here focus on the \textsc{BPASS} grids as they include the effects of massive stars and binary evolution, which are likely more representative of the young stellar populations in our galaxies. We also include the corresponding \textsc{BC03/MILES} grids in Appendix~\ref{app:bc03}, which show qualitatively similar trends.

\begin{figure*}[!ht]
  \centering
\includegraphics[width=\textwidth]{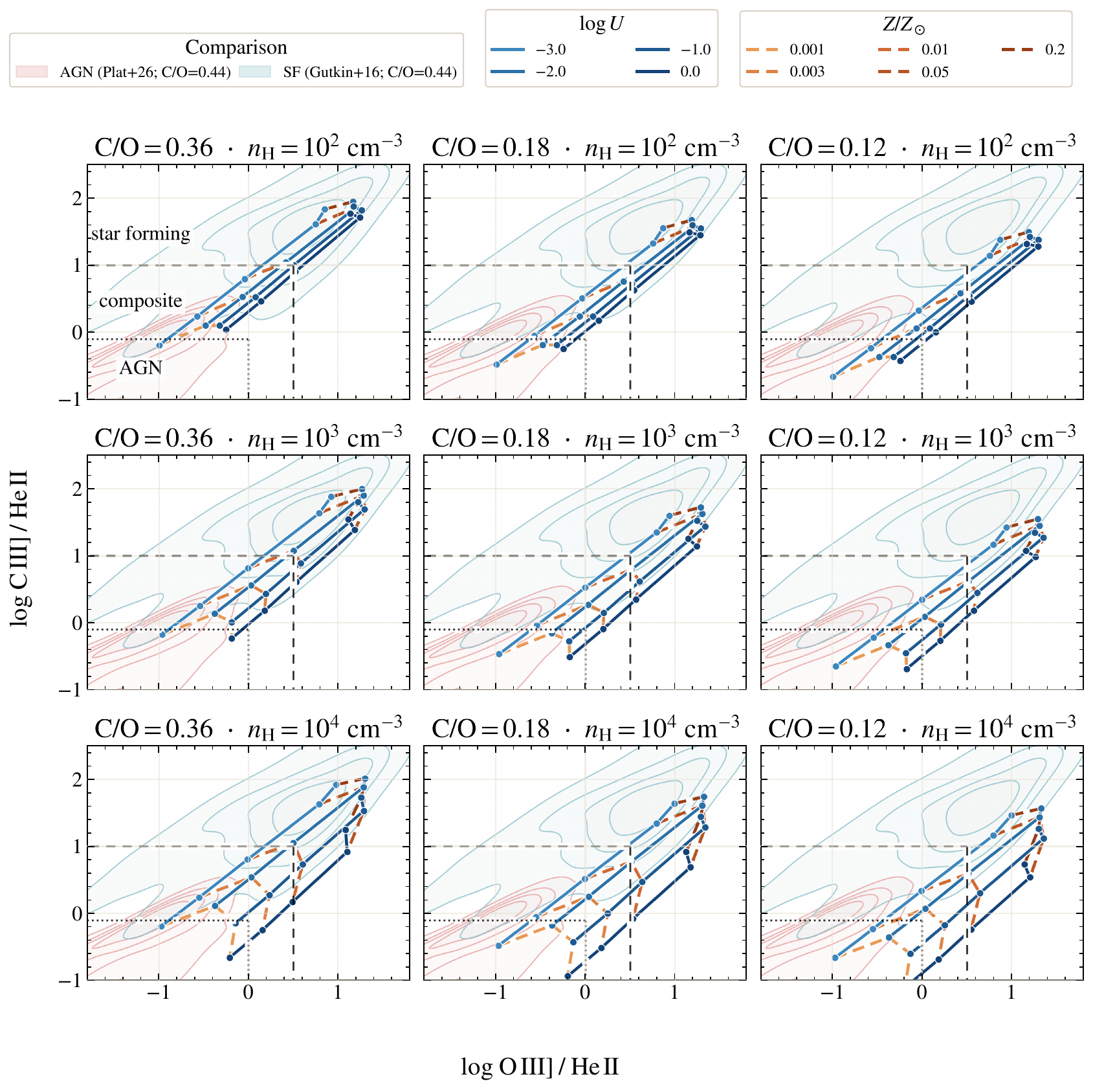}
  \caption{$\log(\oiii]\lambda1666/\heii\lambda1640)$ versus $\log(\ciii\lambda\lambda1907,1909/\heii\lambda1640)$ UV line diagnostic plot.   In each panel, the thick lines show Cloudy/\textsc{BPASS} stellar photoionization model grids.  Each row of panels shows the models at fixed density, $n_{\rm H}=10^2$ (top row), $10^3$ (middle), and $10^4\,\mathrm{cm^{-3}}$ (bottom).   Each column of panels corresponds to models with fixed C/O ratios of 0.36 (left), 0.18 (center), and 0.12 (right). The solid blue lines connect models at fixed $\log U=-3$, $-2$, $-1$, and $0$, with darker shades indicating higher $\log U$ (see legend).  The dashed yellow-to-red lines connect models at fixed $Z/Z_\odot=0.001$, 0.003, 0.01, 0.05,  and 0.2, with redder colors indicating higher metallicity (see legend). Small blue circles mark individual grid points. The blue contours and shading show the \citet{Gutkin2016} predicted star-forming locus; the pink contours and shading show the \citet{Plat2026} predicted AGN narrow-line-region locus. Black dashed and dotted lines mark the \citet{Hirschmann2023} star-forming/composite/AGN boundaries, as labeled in the top-left panel. Increasing the gas density shifts models down and to the right, where the shift increases for higher $\log U$.  Decreasing the C/O shifts the models to lower $\log \ciii/\heii$.}
  \label{fig:uv_o3_loci}
\end{figure*}

The $\log(\oiii]/\heii)$ versus $\log(\ciii/\heii)$ plane is one of the most widely used UV diagnostics for distinguishing star formation from AGN. As shown in Figure~\ref{fig:uv_o3_loci}, for this diagnostic the \citet{Plat2026} AGN locus occupies the lower $\ciii/\heii$ and lower $\oiii]/\heii$ values, as opposed to the \citet{Gutkin2016} star-forming locus, which occupies upper values of these line ratios. F igure~\ref{fig:uv_o3_loci} shows that our BPASS star-forming grids with C/O$=0.36$ and $n_{\rm H}=100~\mathrm{cm^{-3}}$ closely overlap the star-forming locus of \citet{Gutkin2016}. These grids reach the far-composite and AGN regions only at low ionization parameters ($\log U\leq-3.0$) and very low metallicities ($Z/Z_\odot\leq0.001$).   However, reducing the C/O ratio and/or increasing $n_{\rm H}$ moves the center of our grid of $\log(\oiii]/\heii)$ versus $\log(\ciii/\heii)$  towards the composite/AGN region defined by \citet{Hirschmann2023}. This is because lowering C/O reduces the total carbon abundance, suppressing \ciii\ relative to \heii, reducing \ciii/\heii. Increasing the gas density changes the ionization and thermal structure of the nebula, altering the relative strengths of the metal lines and \heii.  Therefore, the interpretation of these UV emission lines depends on the model assumptions of C/O and $n_H$. 

\begin{figure*}[!ht]
  \centering
  \includegraphics[width=\textwidth]{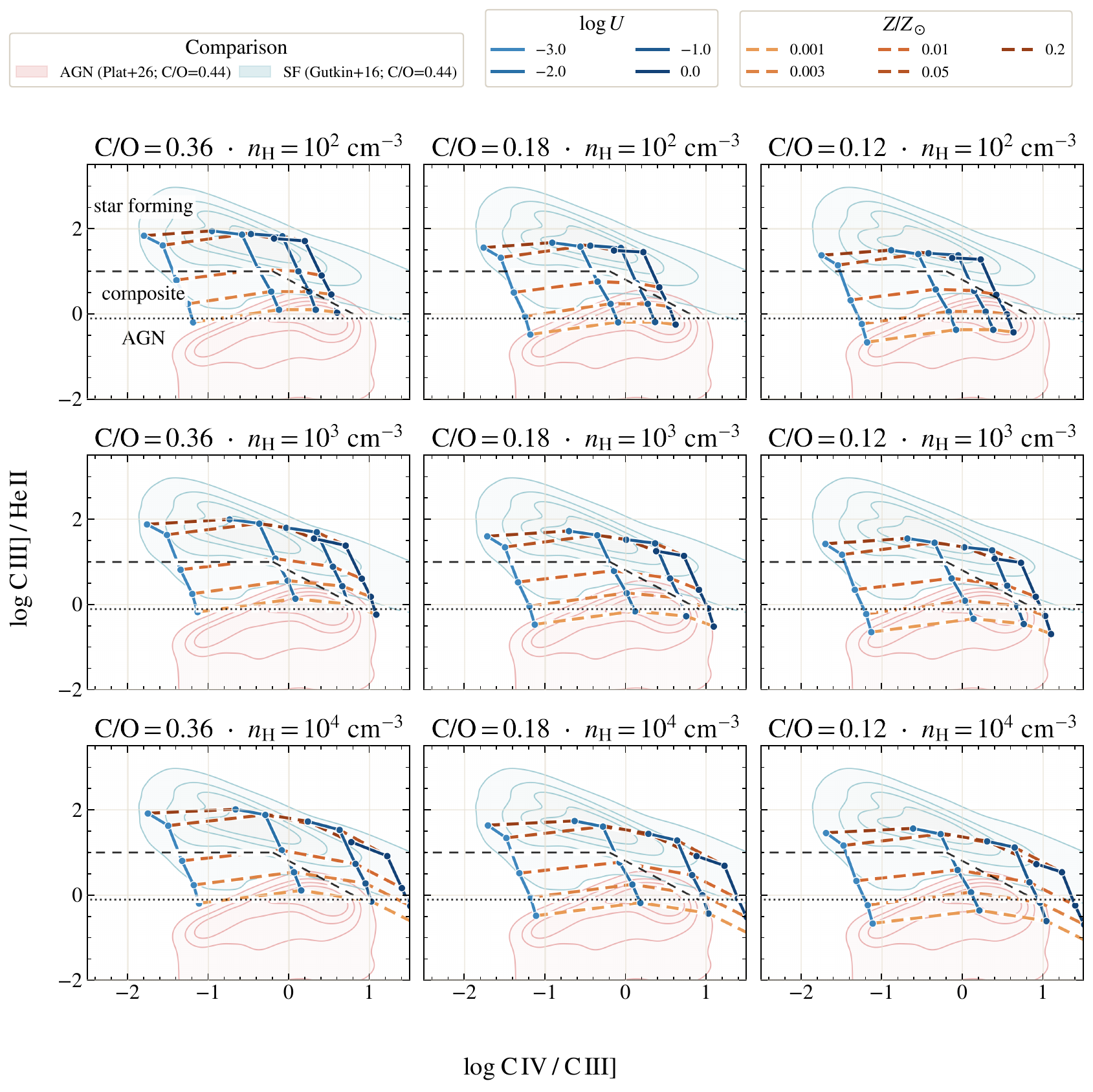}
  \caption{
  $\log(\civ\lambda\lambda1548,1551 / \ciii\lambda1907,1909)$ versus $\log(\ciii\lambda\lambda1907,1909/\heii\lambda1640)$ UV line diagnostic plot.   In each panel, the thick lines show Cloudy/\textsc{BPASS} stellar photoionization model grids.  Each row of panels shows the models at fixed density, $n_{\rm H}=10^2$ (top row), $10^3$ (middle), and $10^4\,\mathrm{cm^{-3}}$ (bottom).   Each column of panels corresponds to models with fixed C/O ratios of 0.36 (left), 0.18 (center), and 0.12 (right). The solid blue lines connect models at fixed $\log U=-3$, $-2$, $-1$, and $0$, with darker shades indicating higher $\log U$ (see legend).  The dashed yellow-to-red lines connect models at fixed $Z/Z_\odot=0.001$, 0.003, 0.01, 0.05,  and 0.2, with redder colors indicating higher metallicity (see legend). Small blue circles mark individual grid points. The blue contours and shading show the \citet{Gutkin2016} predicted star-forming locus; the pink contours and shading show the \citet{Plat2026} predicted AGN narrow-line-region locus. Black dashed and dotted lines mark the \citet{Hirschmann2023} star-forming/composite/AGN boundaries, as labeled in the top-left panel.  Increasing the gas density shifts models to the right, where the shift increases for higher $\log U$.  Decreasing the C/O shifts the models to lower $\log \ciii/\heii$.
}
\label{fig:uv_c43_loci}
\end{figure*}

Another frequently employed diagnostic is the $\civ/\ciii$ ratio, which is sensitive to the hardness of the ionizing spectrum, so it is useful for distinguishing star-forming galaxies from AGN. In this diagnostic, \civ\ denotes the sum of the $\lambda1548$ and $\lambda1551$ fluxes. Figure~\ref{fig:uv_c43_loci} shows that this classification is also sensitive to C/O and density.   Here, increasing $n_{\rm H}$ increases the predicted $\civ/\ciii$ at fixed $\log U$, agreeing with the lower left panel in our heatmap plot (Figure~\ref{fig:line_ratio_heatmap}). This moves the star-forming grid toward and into the parameter space of the \citet{Hirschmann2023} composite/AGN boundaries. Lowering C/O decreases $\ciii/\heii$ while having only a weak impact on $\civ/\ciii$, since both carbon lines scale with the carbon abundance. This produces a downward shift in the diagram.

\begin{figure*}[!ht]
    \centering
    \includegraphics[width=1.0\linewidth]{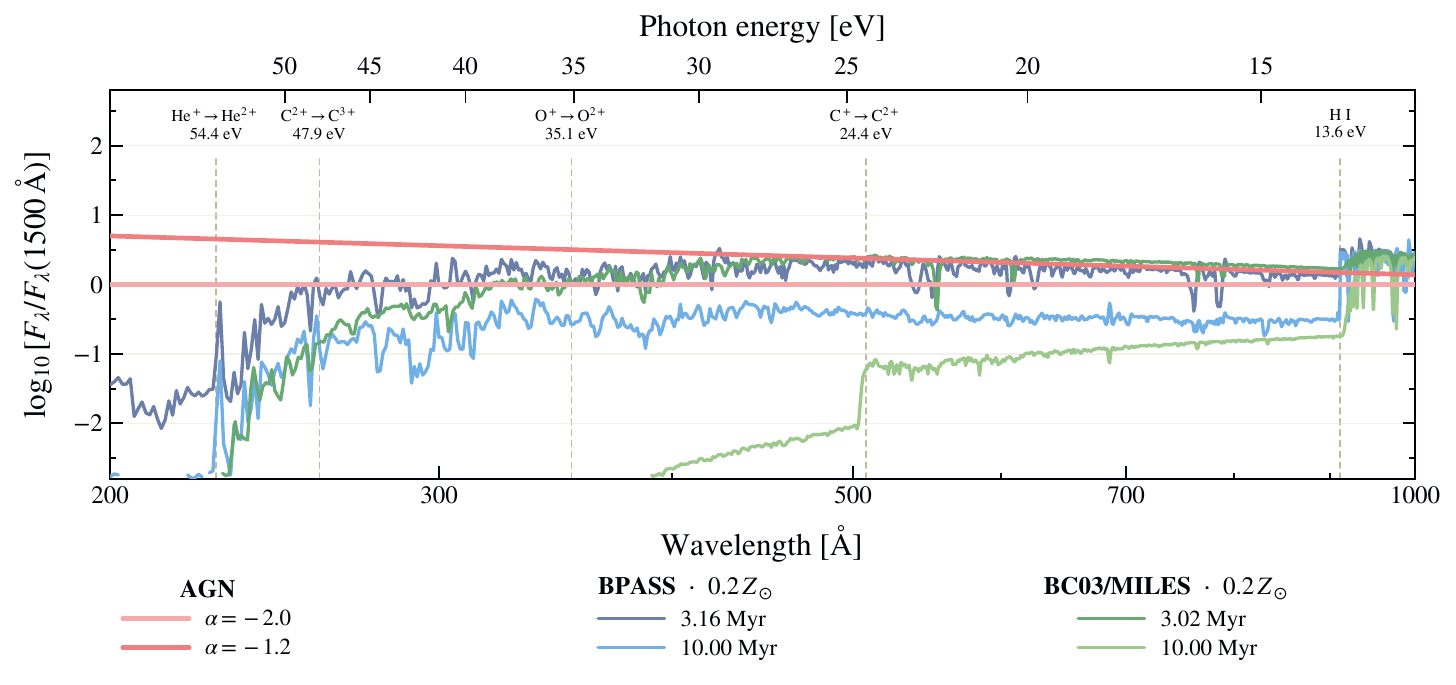}
    \caption{Comparison of the AGN ionizing spectral shapes normalized at $F_\lambda(1500\,\text{\AA})$ to the stellar-population models. The two red curves show AGN power-law continua with $\alpha = -2.0$ and $\alpha = -1.2$, respectively.  The blue and green curves show the \textsc{BPASS} and BC03/MILES models, respectively, for stellar populations with ages of $\approx 3$ and 10 Myr for a metallicity of $0.2Z_{\odot}$ (see legend). Vertical dashed lines mark the ionization thresholds for different ions, as labeled.  The top axis provides the corresponding energy. The figure illustrates how the different spectra produce different ionizing photon rates, $Q(>13.6~\mathrm{eV})$, particularly for higher ionizing lines like \heii.}
    \label{fig:incident_spec}
\end{figure*}

\subsection{AGN Models}
\label{sec: agn}
Motivated by the results above, we also consider the effects of increasing the gas density and changing the C/O ratio for AGN models.  For example, we may expect that reducing the C/O ratio for AGN models will similarly lower \ciii/\heii.   We generate a grid of AGN photoionization models using Cloudy covering the parameter space described in Table~\ref{tab:agn_grid}, which spans a range of C/O, $n_{\rm H}$, $\log U$, metallicity ($Z$), and AGN ionizing continuum slope ($\alpha$).  The models used in the comparisons below are grain-free. We describe the incident AGN photoionization model using Cloudy's power-law prescription with $F_\nu \propto \nu^\alpha$, where we consider $\alpha = -2.0$ and $-1.2$, corresponding to the soft and hard endpoints of the range explored in \citet{Feltre2016}. These values sample the effect of continuum hardness on the predicted emission-line ratios \citep[see][who explore a wider range of physical conditions in AGN]{Plat2026}. Figure~\ref{fig:incident_spec} compares our AGN-model continua with the BPASS and BC03/MILES stellar population spectra. At the same 1500 \AA\ normalization, the AGN continua have more flux above the high-energy ionization thresholds shown in the figure.

\begin{table}
\centering
\begin{tabular}{ll}
\hline
Parameter & Adopted values \\
\hline
C/O & 0.12, 0.18, 0.36 \\
$\log(n_{\rm H}/{\rm cm^{-3}})$ & 2.0, 3.0, 4.0 \\
$\log U$ & $-4.0$, $-3.0$, $-2.0$, $-1.0$ \\
$\alpha$ & $-2.0$, $-1.2$ \\
Metallicity, $Z/Z_\odot$ & 0.001, 0.005, 0.02, 0.06, 0.14, 0.20, 0.40 \\
\hline
\end{tabular}
\caption{Adopted parameters of the grain-free AGN photoionization model grid.}

\label{tab:agn_grid}
\end{table}

\section{Results}
\label{sec:results}
In this section, we compare the photoionization models described in Section~\ref{sec:models} with the observed UV line ratios and equivalent widths for the galaxies both at low and high redshifts.

\subsection{Comparison to Star-Forming Models}
\label{sec:results_sf}
\begin{figure*}[!ht]
  \centering
  \includegraphics[width=\textwidth]{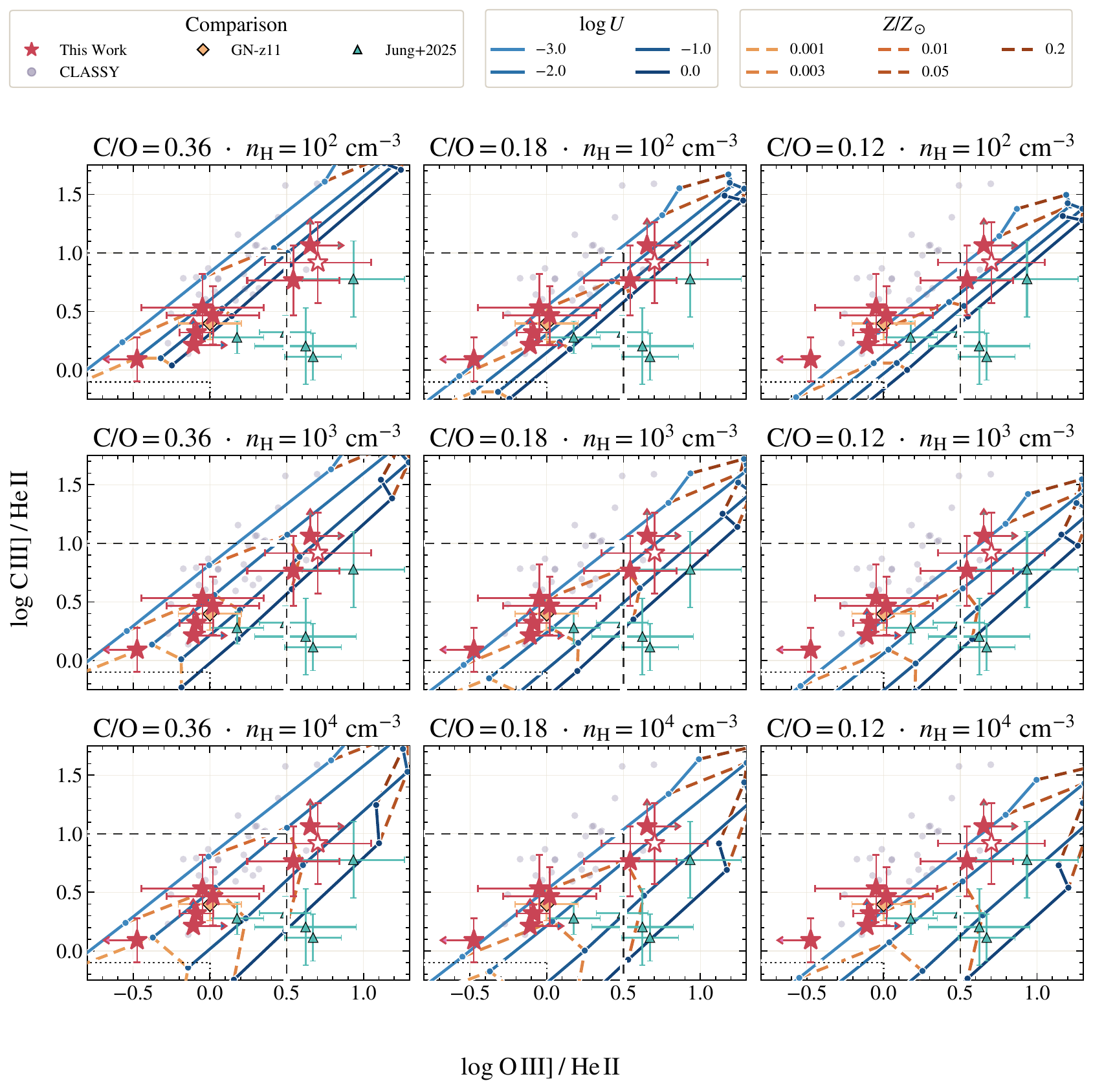}
  \caption{Same as Figure~\ref{fig:uv_o3_loci}, but adding the observed line ratios, and zoomed in on a narrow region of $\log \oiii]  / \heii$ and $\log \ciii/\heii$.  
  The data points show measurements from our \textit{JWST}/NIRSpec targets (red stars, with the open red star marking the LRD; arrows indicate $1\sigma$ upper or lower limits) and literature comparisons, including GN-z11 (orange diamond; \citealt{Bunker2023}), low-redshift sources from \textsc{CLASSY} (pale purple circles; \citealt{Mingozzi2024}) and \citet{Jung2025} (teal triangles).
  }
  \label{fig:uv_o3_data}
\end{figure*}

\begin{figure*}[!ht]
  \centering
  \includegraphics[width=\textwidth]{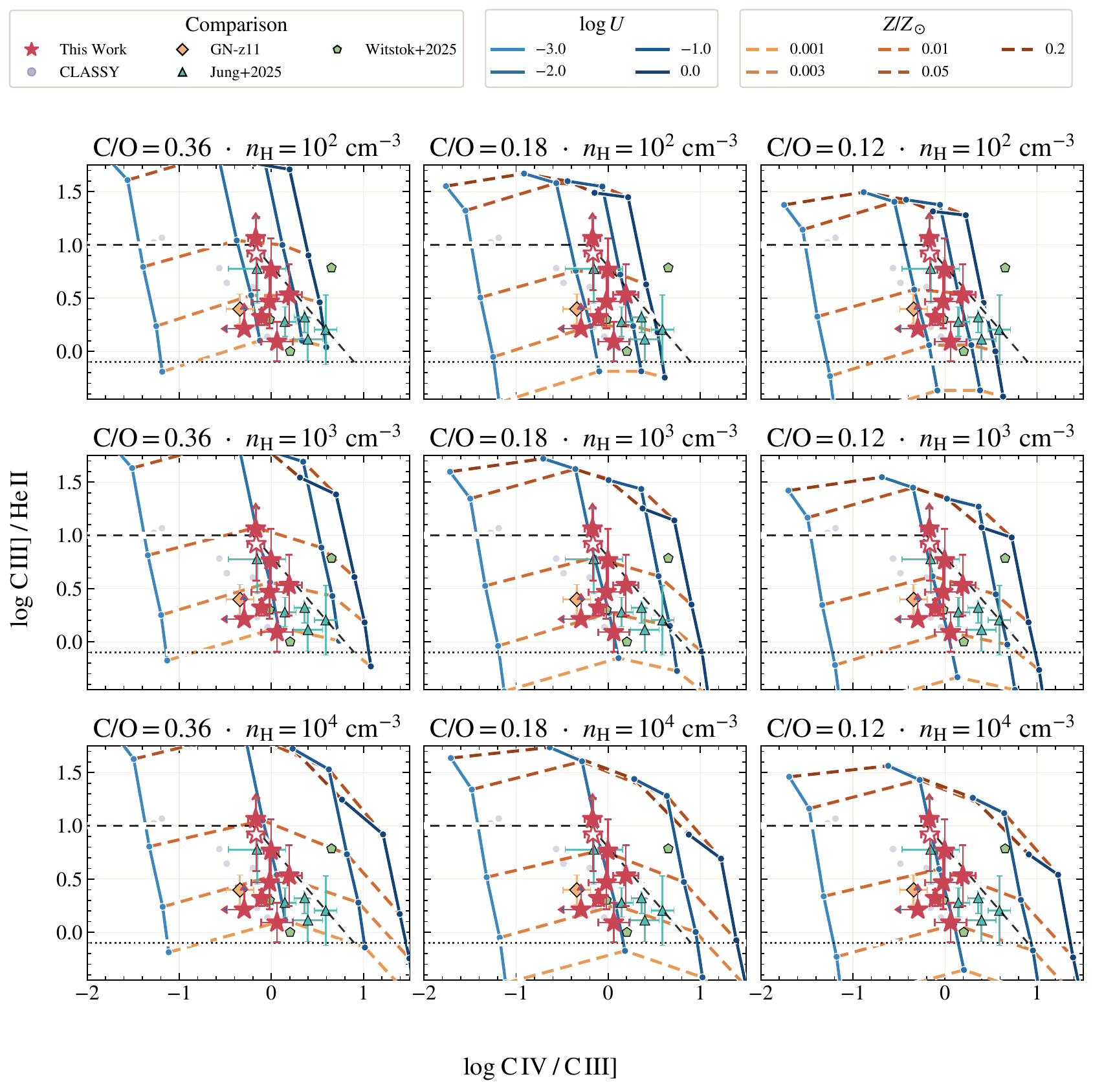}
  \caption{
  Same as Figure~\ref{fig:uv_c43_loci}, but adding the observed line ratios, and zoomed in on a narrow region of $\log \civ\ / \ciii$ and $\log \ciii/\heii$.  
  The data points show measurements from our \textit{JWST}/NIRSpec targets (red stars, with the open red star marking the LRD; arrows indicate $1\sigma$ upper or lower limits) and literature comparisons, including GN-z11 (orange diamond; \citealt{Bunker2023}), low-redshift sources from \textsc{CLASSY} (pale purple circles; \citealt{Mingozzi2024}) and \citet{Jung2025} (teal triangles).
  }
  \label{fig:uv_c43_data}
\end{figure*}

Figures~\ref{fig:uv_o3_data} and \ref{fig:uv_c43_data} compare our new model grids with the \oiii]/\heii, \ciii/\heii, and \civ/\ciii\ emission-line ratios measured for galaxies in our C3PO sample and the literature. Most galaxies in our sample have UV emission-line ratios that would place them in the nominal composite or AGN regions of previous diagnostic diagrams (e.g., \citealt{Hirschmann2023}). However, our new star-forming grids, which allow for lower C/O and higher gas densities, span the observed ratios of galaxies in our sample and the literature. The emission-line ratios of galaxies in our sample are broadly consistent with models having $\log U\sim-2$, in agreement with the values listed in Table~\ref{tab:physical_properties}.

Figure \ref{fig:uv_o3_data} shows that the grids with C/O$\approx 0.12$ and $n_H \approx 10^4~\mathrm{cm^{-3}}$ are broadly consistent with our targets, as well as GN-z11 \citep{Bunker2023}, the local \textsc{CLASSY} analogs \citep{Mingozzi2024}, and the \citet{Jung2025} sample. The AGN-like UV line ratios of these galaxies can therefore be reproduced by purely stellar photoionization within the explored model conditions.

Figure~\ref{fig:uv_c43_data} shows the corresponding comparison in the $\civ/\ciii$--$\ciii/\heii$ plane. Lowering C/O and increasing the gas density shifts the stellar grids towards the observed ratios, consistent with the behavior described in Section~\ref{sec: SF}. Therefore, a position beyond the nominal star-forming boundary can be compatible with stellar photoionization.

The EW$_0$ also provide an important constraint on the physical conditions of the ionized gas in galaxies. Since \ciii\ $\lambda\lambda1907,1909$ is one of the strongest rest-UV metal-line features accessible in these high-redshift galaxies, we here focus EW($\mathrm{C\,III]}$), which is sensitive to the ionization state of the gas, the carbon abundance, metallicity, and nebular density.  Similarly, \ion{O}{3}] is sensitive to the gas density \citep[see, e.g., ][]{Hsiao_2026arXiv260820339H}.  

\begin{figure*}[!ht]
    \centering
    \includegraphics[width=1.0\linewidth]{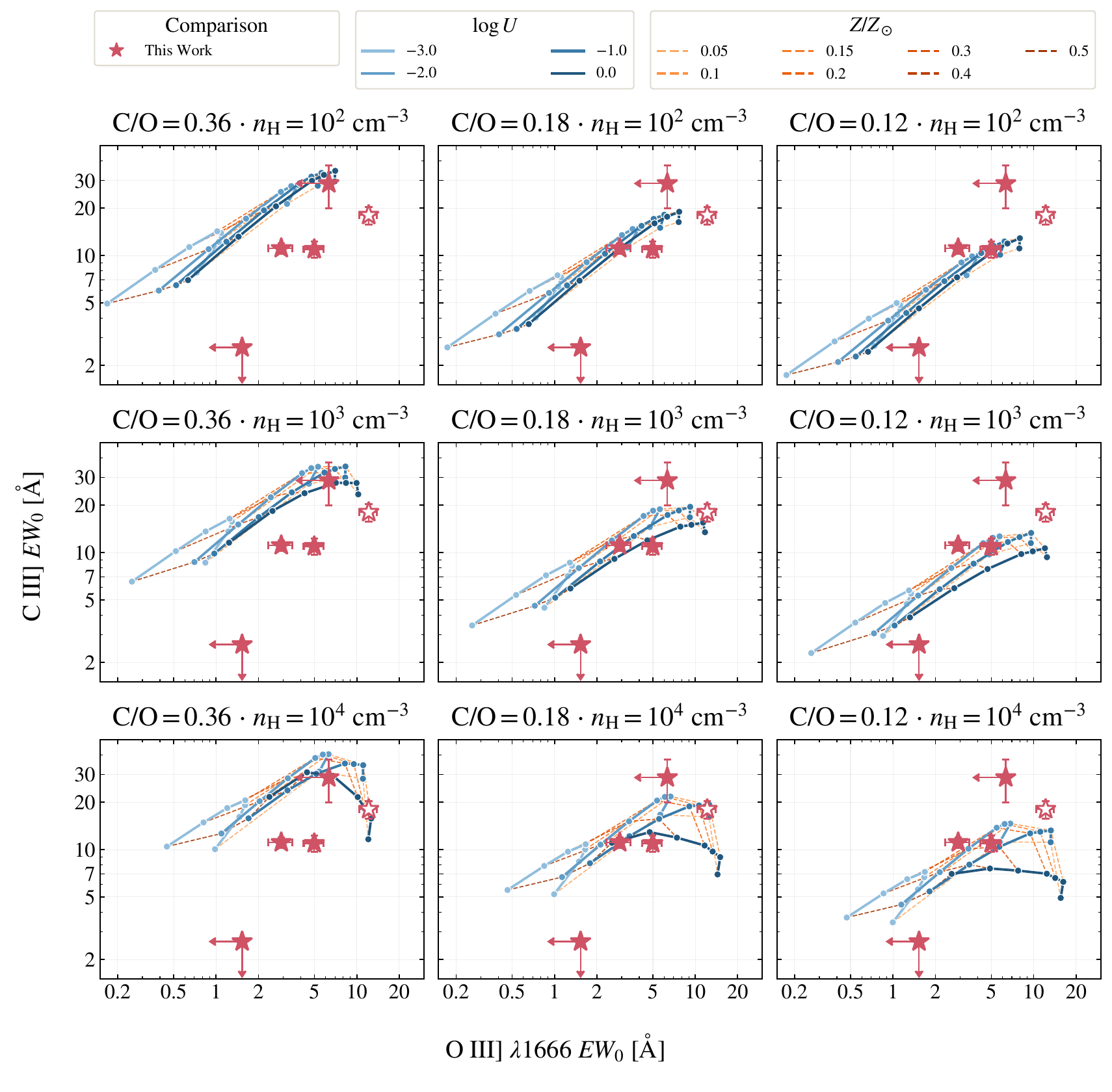}
    \caption{Predictions for rest-frame \ciii\ EW against the rest-frame \ion{O}{3}] EW from the Cloudy/\textsc{BPASS} stellar models. Each row of panels shows the models at fixed density, $n_{\rm H}=10^2$ (top row), $10^3$ (middle), and $10^4\,\mathrm{cm^{-3}}$ (bottom).   Each column of panels corresponds to models with fixed C/O ratios of 0.36 (left), 0.18 (center), and 0.12 (right). The solid blue lines connect models at fixed $\log U=-3$, $-2$, $-1$, and $0$, with darker shades indicating higher $\log U$ (see legend). The dashed orange lines connect models at fixed metallicity, with darker shades indicating higher $Z$ (see legend).   The data points show our targets with robust \ciii\ and \ion{O}{3}] EWs, arrows denote upper limits.  Increasing the density increases the \ion{O}{3}] EW and decreases the predicted \ciii\ EW, especially for higher $\log U$ as the C$^{3+}$ zone increases in size (and the C$^{2+}$ zone shrinks). Decreasing the C/O ratio lowers the \ciii\ EW. Models with higher density and lower C/O better reproduce the data.}
    \label{fig:ew_C3_ew_O3}
\end{figure*}

Figure~\ref{fig:ew_C3_ew_O3} shows the predicted EW$_0$(\ciii) against EW$_0$(\ion{O}{3}]) for our BPASS stellar models and Cloudy results for different values of C/O and $n_e$.   The grid shows the predictions for changing the ionization parameter and metallicity.   Increasing the density increases the \ion{O}{3}] EW slightly because this transition has a higher critical density compared to other lines like \oiii\ \citep[e.g.,][]{Hsiao_2026arXiv260820339H}.  In contrast, increasing the density decreases the predicted \ciii\ EW, especially for higher $\log U$, as the C$^{3+}$ zone increases in size (and the C$^{2+}$ zone shrinks). Decreasing the C/O ratio also lowers the \ciii\ EW. Models with the fiducial solar, C/O=0.36, values can span the range of EW$_0$, but they do not simultaneously reproduce both the EW(\ciii) and EW(\ion{O}{3}]) measured for the galaxies in our sample (save possibly the LRD C3PO 45290, but see \citealt{Papovich2026}).   Models with higher density and lower C/O better reproduce the data.  

\subsection{Comparison to AGN Models}

We have shown that lowering C/O ratios shifts the star-forming galaxy grids of the BPASS model into parts of the composite region, with more limited overlap with the AGN-dominated region defined by commonly used rest-UV diagnostic diagrams \citep{Feltre2016, Gutkin2016, Hirschmann2023, Plat2026}. Here, we compare the observations with the grain-free AGN models described in Section~\ref{sec: agn} and probe their dependence on C/O and gas density.

\begin{figure*}[p]
    \centering
    \includegraphics[width = 0.97\linewidth]{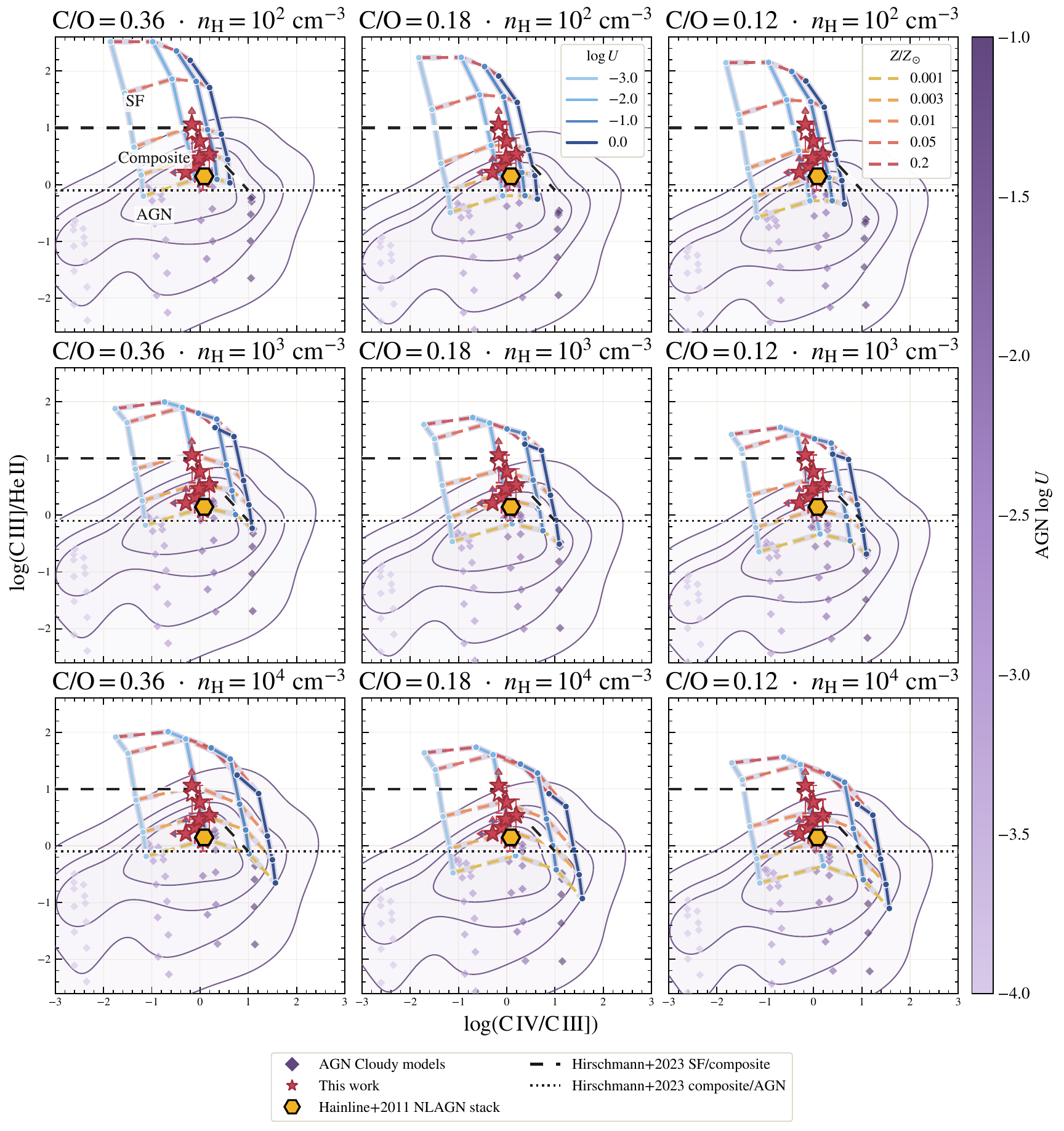}
    \caption{Comparison between AGN photoionization models and \textsc{BPASS} stellar photoionization grids in the rest-UV $\civ/\ciii$--$\ciii/\heii$ diagnostic plane. In each panel,  colored diamonds show individual grain-free AGN models spanning ionization parameter, metallicity, and continuum slope (Table \ref{tab:agn_grid}), with color indicating AGN \(\log U\). Purple contours show the smoothed distribution of the plotted AGN models.  Each row of panels shows the models at fixed density, $n_{\rm H}=10^2$ (top row), $10^3$ (middle), and $10^4\,\mathrm{cm^{-3}}$ (bottom).   Each column of panels corresponds to models with fixed C/O ratios of 0.36 (left), 0.18 (center), and 0.12 (right). The colored lines show the BPASS-based nebular grids as in Figure~\ref{fig:uv_c43_loci}, with solid tracks connecting models at fixed $\log U$ and dashed tracks connecting models at fixed metallicity. The dashed and dotted black lines show the \citet{Hirschmann2023} boundaries separating the star-forming, composite, and AGN regions. The gold hexagon shows the stacked rest-UV spectrum of $z\sim2$ narrow-line AGNs from \citet{Hainline2011}. Red stars show the observed JWST sample with measurement uncertainties, with the open red stars showing the LRD.}
    \label{fig:agn}
\end{figure*}

In Figure~\ref{fig:agn}, we compare our new AGN models with our \textsc{BPASS} stellar grids in the $\civ/\ciii$--$\ciii/\heii$ plane, evaluated at each nominal C/O and $n_{\rm H}$. As C/O decreases, \textit{both} the AGN models and the stellar models shift toward lower \ciii/\heii.  This is a result of the reduced carbon abundance, while helium remains unchanged. Compared to the observed galaxies, the measurements are more consistent with the stellar models with lower C/O, while AGN models shift away from the data points.   Decreasing the C/O abundance has the effect of shifting both the star-forming and AGN models to lower $\ciii/\heii$.    

Increasing the density has less of an effect on the AGN models.  The change in density is primarily seen as it shifts the stellar-model grids further into the ``composite'' region of Figure~\ref{fig:agn}, primarily by increasing the predicted $\civ/\ciii$ ratio for higher density.   At higher densities and lower C/O, the stellar-model grids overlap more with the AGN-model grids. Therefore, a location in the nominal composite region does not, by itself, require an AGN contribution.

\begin{figure*}[p]
    \centering
    \includegraphics[width = 0.97\linewidth]{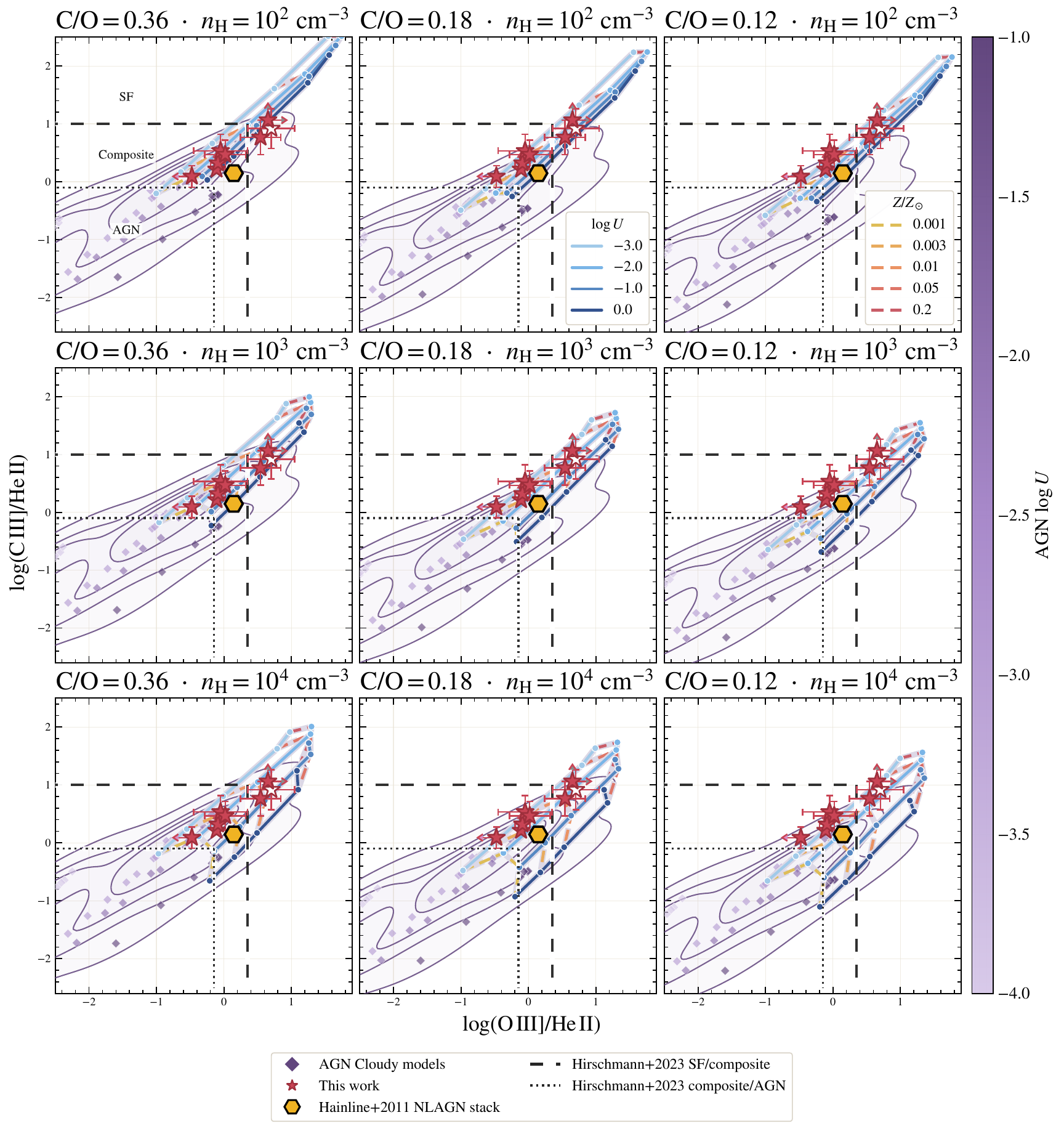}
    \caption{Comparison between AGN photoionization models and BPASS stellar photoionization grids in the rest-UV $\mathrm{OIII]}/\mathrm{HeII}$--$\mathrm{CIII]}/\mathrm{HeII}$ diagnostic plane. In each panel,  colored diamonds show individual grain-free AGN models spanning ionization parameter, metallicity, and continuum slope (Table \ref{tab:agn_grid}), with color indicating AGN \(\log U\). Purple contours show the smoothed distribution of the plotted AGN models.  Each row of panels shows the models at fixed density, $n_{\rm H}=10^2$ (top row), $10^3$ (middle), and $10^4\,\mathrm{cm^{-3}}$ (bottom).   Each column of panels corresponds to models with fixed C/O ratios of 0.36 (left), 0.18 (center), and 0.12 (right). The colored lines show the BPASS-based nebular grids as in Figure~\ref{fig:uv_o3_loci}, with solid tracks connecting models at fixed $\log U$ and dashed tracks connecting models at fixed metallicity. The dashed and dotted black lines show the \citet{Hirschmann2023} boundaries separating the star-forming, composite, and AGN regions. The gold hexagon shows the stacked rest-UV spectrum of $z\sim2$ narrow-line AGNs from \citet{Hainline2011}. Red stars show the observed JWST sample with measurement uncertainties, with the open red stars showing the LRD.}
    \label{fig:agn_bpass_oiii_ciii}
\end{figure*}

Figure~\ref{fig:agn_bpass_oiii_ciii} shows a comparison of the AGN and stellar models in the \oiii]/\heii--\ciii/\heii\ plane. We use this diagnostic because it directly probes the relative strength of carbon and oxygen.  
Decreasing C/O shifts both the AGN and BPASS stellar grids toward lower \ciii/\heii, with only small changes in \oiii]/\heii. Across the explored parameter space, the observed line ratios of our C3PO sample are broadly consistent with stellar photoionization models. As in Figure~\ref{fig:agn}, a location in the composite region does not, by itself, distinguish stellar photoionization from an AGN contribution.
 
These results suggest that comparisons restricted to local fiducial C/O ratios may not fully capture the conditions in the reionization era galaxies, and that low-C/O model subsets warrant consideration when classifying high-redshift sources in rest-UV diagrams.

\section{Discussion}
\label{sec:discussion}

A major finding of this work is that lower C/O ratios and higher gas densities in stellar photoionization models can better reproduce the observed rest-frame UV emission line equivalent widths and line ratios for galaxies in the EoR.  Here, we discuss detailed comparisons between the measured lines and the models. 

\subsection{Interpreting UV Diagnostics at High Redshift}

Our comparisons show that the interpretation of rest-UV diagnostic diagrams depends on the adopted abundance patterns, gas density, and ionizing spectra. Changes in these assumptions, particularly the C/O abundance and gas density, have the effect of shifting predictions for UV emission lines from stellar photoionization models such that they can occupy regions of UV emission-line diagnostics typically associated with composite sources and, to a lesser extent, AGN.  Because EoR galaxies appear to favor lower C/O and higher gas densities, these conditions should be considered when interpreting their locations in UV emission-line diagnostic diagrams.

The star-forming and AGN grids of \citet{Gutkin2016}, \citet{Feltre2016}, \citet{Hirschmann2019}, and \citet{Plat2026} provide an excellent foundation for these comparisons. Their parameter spaces already include low metallicities and high gas densities, with the star-forming models also exploring variations in C/O. Subsequent work connected these models to simulated galaxy populations and developed classification boundaries for UV diagnostic diagrams that successfully interpret the conditions in galaxies at low and moderate redshifts \citep{Gutkin2016, Hirschmann2017, Hirschmann2019, Hirschmann2023}.  These models include a wide range of parameter space, including lower C/O conditions (0.1 to 1.0 times solar), and gas densities of $10^2$ and $10^3$ cm$^{-3}$.  When restricting these predictions to lower C/O values, they shift in a similar way to what we observe here \citep[e.g., Figure~10 of ][and discussion in \citealt{Plat2026}]{Papovich2026}.   Therefore, these models are valuable, but they need to be restricted to the parameter space best matched to observed conditions in EoR galaxies. 

Our measured metallicities and C/O ratios motivate comparisons with metal-poor, sub-solar-C/O models (Figure \ref{fig:co_oh_relation}). Our measurements of the electron densities from \ciii\ (see above) are consistent with other JWST measurements that indicate elevated nebular electron densities in high-redshift galaxies relative to nearby samples using \oii, \ciii, and \oiii-based measures \citep{Isobe2023,Martinez_2025,Topping2025,Hsiao_2026arXiv260820339H}. Density comparisons require care however, because the different lines probe different ionization zones. In particular, the electron density inferred from \ciii\ characterizes the carbon-emitting gas and should not be identified directly with the uniform hydrogen density adopted in photoionization models. For example, \citet{Hsiao_2026arXiv260820339H} recently demonstrated that gas densities derived from \oiii\ emission line ratios in some high-redshift galaxies favor values as high as $10^5 - 10^6$~cm$^{-3}$ in the O$^{2+}$ zone, substantially higher than measured from lower-ionization lines (e.g., $n_e([$\ion{S}{2}$])$) for high-redshift galaxies \citep{Reddy_2023}. These observations motivate exploring how elevated density affects the predicted UV line ratios.

Building on earlier studies, we use these observational constraints to guide our comparisons with stellar and AGN photoionization grids. The overlap between stellar models and parts of the nominal composite or AGN regions illustrates how abundance and density assumptions can affect classification. The observed ratios remain compatible with stellar photoionization within the explored parameter space, although this agreement does not exclude an AGN contribution. Our results, therefore, support evaluating the measured line ratios against models informed by the physical conditions of the sample, rather than assigning an ionizing source from a classification boundary alone.
A complementary limitation arises on the AGN side. \citet{Plat2026} find that several classical diagnostic diagrams can miss low-metallicity AGN. Their narrow-line calculations show that metallicity and $\log U$ influence the predicted ratios more strongly than black-hole mass or Eddington ratio.

The adopted stellar population introduces an additional dependence. Binary evolution can prolong the production of hard ionizing photons, providing a physical motivation for our use of BPASS \citep{Eldridge2017, Stanway2018}. The BPASS and BC03/MILES comparisons presented in the Appendix show similar qualitative shifts across the diagnostic diagrams, but do not establish that either model uniquely describes these galaxies. The interpretation remains conditional on the adopted stellar age, star-formation history, and nebular assumptions.

\subsection{Deficit of High-Energy Photons}

\begin{figure}
    \centering
    \includegraphics[width=\linewidth]{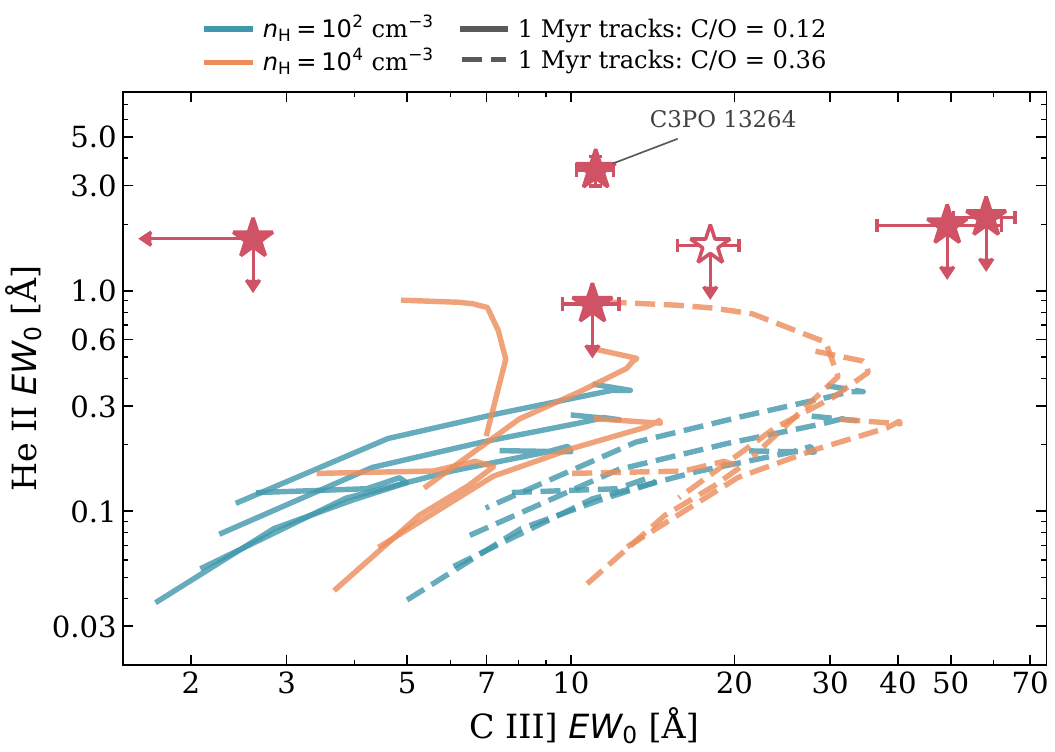}
    \caption{Predictions for rest-frame \heii\ EW against rest-frame \ciii\ EW from the Cloudy/\textsc{BPASS} stellar models. The teal and orange lines show models with gas densities $n_{\rm H}=10^2$ and $10^4\,\mathrm{cm^{-3}}$, respectively. Solid and dashed lines correspond to C/O ratios of 0.12 and 0.36, respectively (see legend). The data points show our sample, with an open star marking the LRD C3PO-45290. Error bars show $1\sigma$ uncertainties. Arrows denote $1\sigma$ EW upper limits, with downward arrows for \heii\ lines with $\mathrm{S/N}<2$ and leftward arrows for \ciii.}
    \label{fig:heii_ciii_ew}
\end{figure}

It is noteworthy that our models used here underpredict the strength of \heii\ EW in at least one galaxy in our sample, C3PO-13264.  Figure~\ref{fig:heii_ciii_ew} shows that models of a burst with age 1\,Myr and gas density $n_e = 10^2$~cm$^{-3}$ produce a peak \heii\ EW$_0$ up to $\simeq0.4$\,\AA.  This increases the peak \heii\ EW$_0$ to $\approx 1.2$\,\AA\ for a gas density of $n_e = 10^4$~cm$^{-3}$.  Changing the age of the burst does not increase the \heii\ EW$_0$, nor does varying the metallicity or C/O further. Within the models explored here, changing the age of the burst does not increase the \heii\ EW$_0$, nor does varying the metallicity or C/O further. In contrast, at least one galaxy in our sample, C3PO 13264, has a \heii\ EW$_0$ above this value, with EW$_0=3.53\pm0.55$\,\AA\ (other galaxies have limits, see Table~\ref{tab:linefluxes} and Figure~\ref{fig:heii_ciii_ew}. This implies our models produce insufficient high-energy photons with $E > 54.4$~eV relative to the UV continuum. 

This tension is not unique to the reionization era. \citet{Steidel2016} showed that standard single-star population-synthesis models struggle to reproduce the high-excitation nebular spectra of typical $z \simeq 2.4$ star-forming galaxies. \citet{Olivier2022} found that reproducing the high-ionization lines of local extreme emission-line galaxies requires an additional hot component---modeled as an $\simeq 80{,}000$ K blackbody that supplies a substantial fraction of ionizing luminosity---while preserving the low-ionization constraint. Similar hard-photon deficits have also been reported in metal-poor systems both among high-z galaxies and their local analogs \citep{Berg2018, Senchyna2023} or as evidence for very massive stars (VMS, \citealt{Upadhyaya_2024A&A...686A.185U}).
Although difficulties remain for some stellar-population prescriptions \citep{Eldridge2022}, alternative treatments of binary evolution can produce sufficient hard ionizing photons for strong nebular \heii\ emission \citep{Lecroq2024}.

Several sources could supply the missing high-energy photons. The high \heii\ EW$_0$ could indicate the presence of an AGN, but this would also be expected to impact additional high-energy lines such as [\ion{Ne}{5}]$\lambda3427$ \citep[e.g.,][]{Cleri_2023}, which remains undetected in our data.  Hot Wolf--Rayet stars and stars stripped in binaries are plausible sources of He$^{+}$-ionizing photons \citep{Gotberg2019}. VMS ($M_{\rm init}>100\,M_\odot$) can also contribute directly to the line through stellar winds. Models including these stars reproduce the broad \heii\ and UV wind features observed in two galaxies at $z\simeq8.7$ \citep{Marques-Chaves2026} and in 13 galaxies at $z\sim 2-3$ \citep{Upadhyaya_2024A&A...686A.185U}. Their contribution to nebular \heii\ depends on how age, metallicity, and mass loss affect the ionizing flux emerging from their winds \citep{Martins2025}. Accreting compact objects could also supply hard photons \citep{Richardson2025}, although the X-ray luminosities and upper limits measured by \citet{Senchyna2020} were too low for their high-mass X-ray binary models to explain the nebular \heii\ in their nearby metal-poor sample.

Our next step is to test whether adding these other sources of high-energy photons to the stellar models can reproduce the \heii\ EW$_0$ and other high-ionization lines, while preserving the measurements of \ion{O}{3}] and  \ciii\ EW. Complementary diagnostics based on other emission line ratios and equivalent width 
%
%
could help distinguish the effects of low C/O from an AGN contribution \citep[see, e.g.,][]{Plat2026}.  This could include programs that target \ion{Ne}{5}, which requires photons above 97~eV, and also \neiv$\lambda2425$, which requires photons above $63.4$\,eV. This line has already been used as a diagnostic in high-z galaxies \citep[e.g.,][]{Brinchmann2023}. Measurements of these lines, combined with \heii, together would probe the shape of the ionizing spectrum beyond the He$^{+}$ edge.  These kinds of studies are crucial to carefully determine the shape of the ionizing spectrum at high energies and the nature of the ionizing sources. 

\section{Conclusion}
\label{sec:conclusion}

We have presented stellar photoionization model grids for rest-frame UV emission-line diagnostics and compared them with JWST/NIRSpec G140M and G395M spectra of eight C3PO galaxies at $z\sim7$--9.   The models allow both for lower C/O ratios (spanning 1/3 solar to solar) and higher gas densities, ranging from $n_e = 10^2$ to $10^4$~cm$^{-3}$.  From the C3PO data, we have measured emission-line fluxes and equivalent widths, or upper limits, for the key rest-UV and optical lines (Table~\ref{tab:linefluxes}).  Our main conclusions are as follows. 

\begin{enumerate}

\item Using \textsc{PyNeb} and the available temperature- and density-sensitive diagnostics, we derive the nebular physical conditions and abundances of eight selected C3PO targets summarized in Table~\ref{tab:physical_properties}. For these EoR galaxies, the measured oxygen abundances are approximately $12+\log(\mathrm{O/H})=7.6$--7.9, well below the solar abundance. We infer the sub-solar C/O ratios spanning from 0.11 to 0.32. The constrained \ciii-based electron densities span approximately $n_{\rm e}\sim10^3$--$5\times10^4\,\mathrm{cm^{-3}}$.  These values are consistent with other literature measurements of high-redshift galaxies.   Measurements with weak lines or adopted fiducial temperatures and densities remain conditional on these assumptions. These physical property measurements motivate the construction of metal-poor, sub-solar-C/O, higher-gas-density photoionization models.

\item We construct photoionization grids using \textsc{Cloudy} C25.00 with \textsc{BPASS} v2.2.1 binary and BC03/MILES stellar populations. The grids explore $\log U=-4$ to 0, C/O ratios of 0.12, 0.18, and 0.36 (adopted as the solar value), and hydrogen densities of $n_{\rm H}=10^2$, $10^3$, and $10^4\,\mathrm{cm^{-3}}$. They provide a framework for examining how abundance patterns, gas density, and the adopted stellar population affect the interpretation of UV emission-line ratios.

\item Lowering the C/O value reduces the predicted carbon-line emission relative to \heii, shifting the model grids toward lower \ciii/\heii\ and \civ/\heii. Increasing the gas density further modifies the carbon-line ratios and the locus of the model grids. Together, the pure star-forming models can extend to occupy parts of the nominal composite and AGN regions defined by commonly used UV diagnostic boundaries assuming lower gas densities and higher C/O values.  We also explore AGN models with lower C/O values, and find these models similarly shift to lower \ciii/\heii\ and \civ/\heii.    Therefore, a galaxy's position in these regions does not necessarily for these UV line diagnostics separate star-forming ionization from AGN contributions. These classifications need to be adjusted based on the metallicity, C/O, and gas density of the galaxies.

\item There is a remaining discrepancy in the \heii\ EW$_0$ for at least one galaxy in our sample, where our stellar population models are unable to reproduce the strong line. Possible solutions to this include the need to add Wolf-Rayet stars, VMS, or AGN. This motivates future studies that will focus on high-ionization lines that probe the shape of the ionizing spectra above the \heii\ edge, $> 54.4$~eV.  This could include such lines as [\ion{Ne}{5}] or \neiv, which, when combined with \heii, \ciii, \ion{O}{3}], etc.  These kinds of studies will be important in order to determine the shape of the ionizing spectrum at high energies and the nature of the ionizing sources during the EoR.

\end{enumerate}

Taken together, these results indicate that the physical conditions characteristic of early galaxies --- chemical youth manifested as sub-solar C/O, and a dense ionized ISM --- are sufficient to account for the rest-UV emission-line ratios of our $z \gtrsim 7$ sample, without necessarily requiring AGN. As JWST continues to build statistical samples of rest-UV spectra at $z>6$, our photoionization grids will be essential for correctly identifying the dominant ionizing sources in reionization-era galaxies.

\begin{acknowledgements}

We thank our colleagues and collaborators for valuable contributions and input that helped direct this work.  In particular, we deeply thank D.~Berg, M.~Hirschmann,  Z.~Martinez,  A.~Plat, and L.~Shen for valuable discussions, feedback, comments, and suggestions.   We also thank A.~Plat and M.~Hirschmann for providing the predictions from their models electronically.   This work benefited from support from the George P. and Cynthia Woods Mitchell Institute for Fundamental Physics and Astronomy at Texas A\&M University. CP thanks Marsha and Ralph Schilling for generous support of this research.  
This work is based on observations made with the NASA/ESA/CSA James Webb Space Telescope. The data were obtained from the Mikulski Archive for Space Telescopes at the Space Telescope Science Institute, which is operated by the Association of Universities for Research in Astronomy, Inc., under NASA contract NAS 5-03127 for JWST. These observations are associated with program \#5943.  Support for US investigators in program \#5943 was provided by NASA through a grant from the Space Telescope Science Institute, which is operated by the Association of Universities for Research in Astronomy, Inc., under NASA contract NAS 5-03127.

    \textit{Facilities:} JWST

    \textit{Software:} Cloudy (C25.00), \textsc{Bagpipes}, JWST Calibration Pipeline v1.20.2 \citep{Bushouse2025}
\end{acknowledgements}

\section*{DATA AVAILABILITY}
The photoionization model grids used in this work will be made publicly available on Zenodo upon publication of this article under the reserved DOI \url{https://doi.org/10.5281/zenodo.23070363}. The associated GitHub repository is available at \url{https://github.com/AnnabellaYang/bagpipes-cloudy25-grids/tree/main}.

\appendix

\section{BC03/MILES Stellar Population Models}
\label{app:bc03}

\begin{figure*}[!ht]
    \centering
    \includegraphics[width=\textwidth]{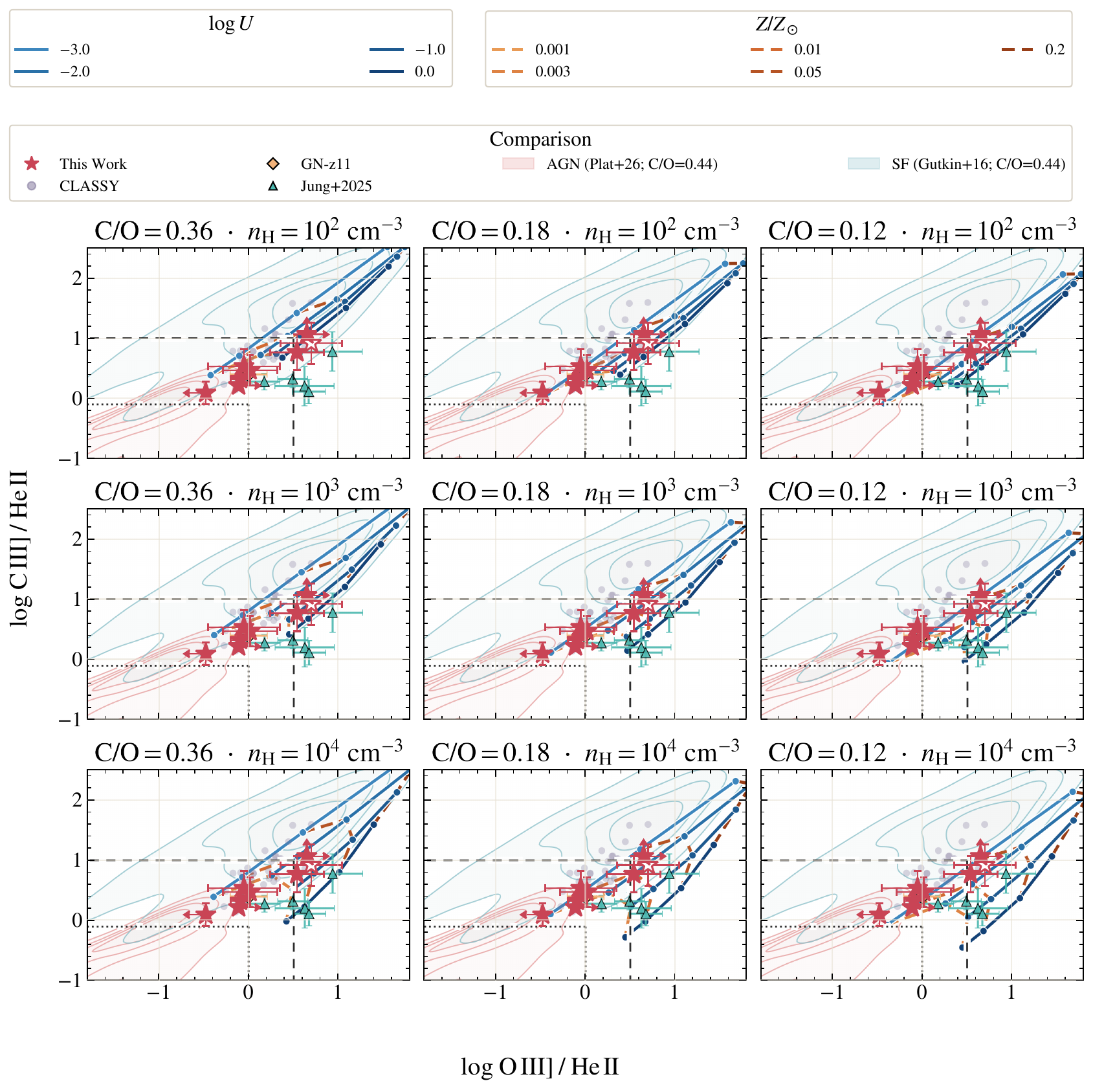}
    \caption{\oiii]/\heii\ versus \ciii/\heii\ diagnostic using BC03/MILES stellar population models. The observed data and comparison loci from Figures~\ref{fig:uv_o3_loci} and \ref{fig:uv_o3_data} are overlaid on the stellar grids. Colours, symbols, and line styles follow those figures.}
    \label{fig:bc03_o3_combined}
\end{figure*}

\begin{figure*}[!ht]
    \centering
    \includegraphics[width=\textwidth]{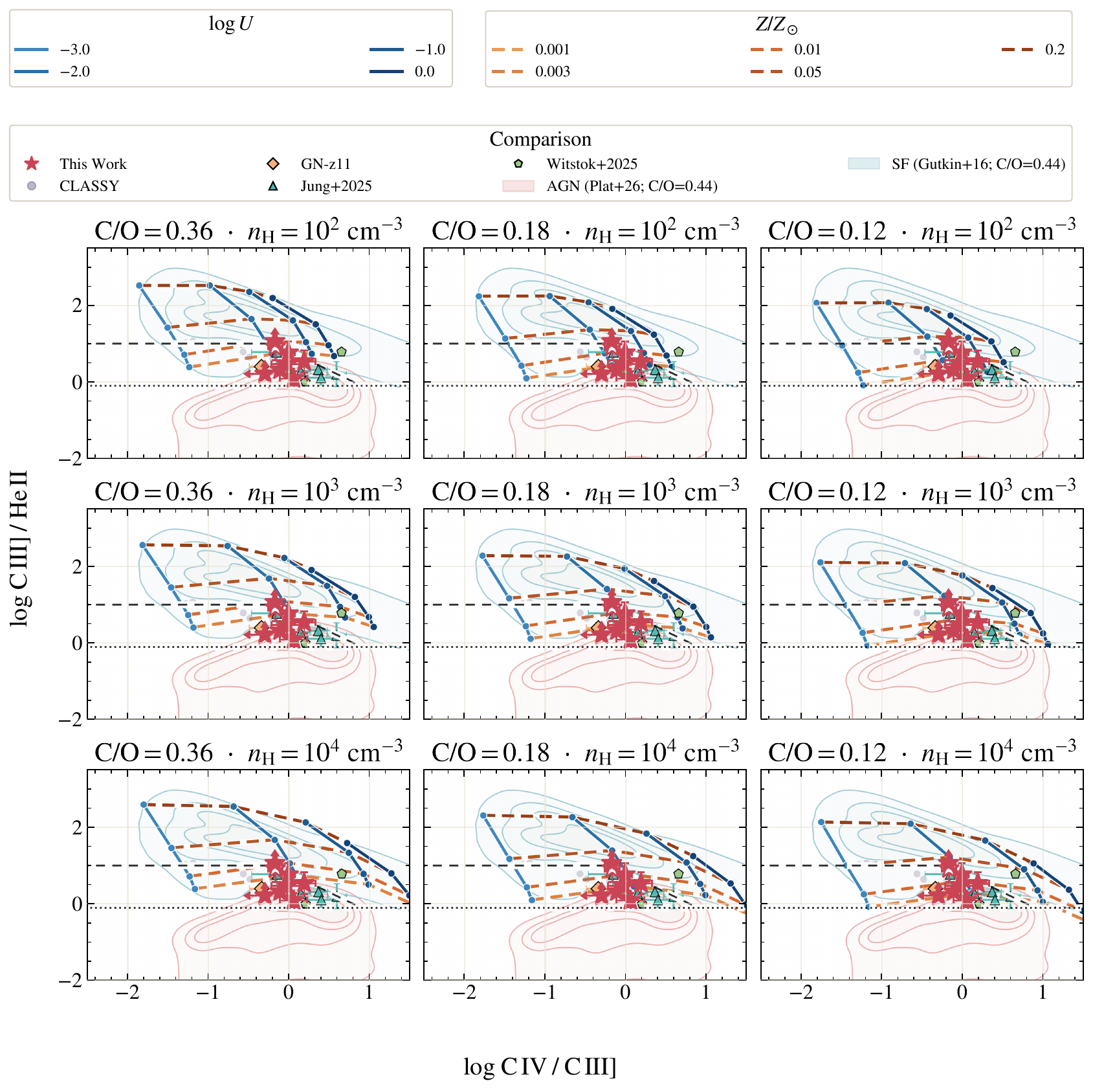}
    \caption{\civ/\ciii\ versus \ciii/\heii\ diagnostic using BC03/MILES stellar population models. The observed data and comparison loci from Figures~\ref{fig:uv_c43_loci} and \ref{fig:uv_c43_data} are overlaid on the stellar grids. Colours, symbols, and line styles follow those figures.}
    \label{fig:bc03_c43_combined}
\end{figure*}

To assess the sensitivity of our results to the choice of stellar population synthesis model, we repeat the diagnostics in Section~\ref{sec:results_sf} using \textsc{BC03/MILES} \citep{BruzualCharlot2003}. Figures~\ref{fig:bc03_o3_combined} and \ref{fig:bc03_c43_combined} show two representative comparisons, with the observed data and model loci overlaid. The other diagnostics tested show qualitatively similar trends to the \textsc{BPASS} case and are omitted for brevity.

The \textsc{BC03/MILES} grids broadly follow the same trends with C/O and gas density as the \textsc{BPASS} grids. However, within the parameter range explored, they do not extend to the lowest \ciii/\heii\ and \oiii]/\heii\ ratios reached by \textsc{BPASS}. Thus, the overall diagnostic behavior is similar, but the extent of the low-ratio model loci depends on the adopted stellar population model. This highlights the importance of improving our understanding of the ionizing radiation for the model stellar populations. 

\clearpage
\bibliography{CO_c3po}{}
\bibliographystyle{aasjournalv7}

\end{document}

%% file: defs.tex
\def\lesssim{\mathrel{\hbox{\rlap{\hbox{%
 \lower4pt\hbox{$\sim$}}}\hbox{$<$}}}}
\def\gtrsim{\mathrel{\hbox{\rlap{\hbox{%
 \lower4pt\hbox{$\sim$}}}\hbox{$>$}}}}
\def\I.{\kern.2em{\sc i}} 
\def\II.{\kern.2em{\sc ii}} 
\def\III.{\kern.2em{\sc iii}} 
\def\IV.{\kern.2em{\sc iv}} 
\def\arcs{\hbox{$^{\prime\prime}$}}

\newcommand{\hst}{\textit{HST}}
\newcommand{\spitzer}{\textit{Spitzer}}
\newcommand{\chandra}{\textit{Chandra}}
\newcommand{\sirtf}{\textit{Spitzer}}
\newcommand{\cxo}{\textit{CXO}}
\newcommand{\hdf}{HDF--N}
\newcommand{\wfu}{\hbox{$U_{300}$}}
\newcommand{\wfb}{\hbox{$B_{450}$}}
\newcommand{\wfv}{\hbox{$V_{606}$}}
\newcommand{\wfi}{\hbox{$I_{814}$}}
\newcommand{\acsv}{\hbox{$V_{606}$}}
\newcommand{\acsi}{\hbox{$I_{814}$}}
\newcommand{\nicj}{\hbox{$J_{110}$}}
\newcommand{\nich}{\hbox{$H_{160}$}}
\newcommand{\mirifive}{\hbox{$[5.6]$}}
\newcommand{\miriseven}{\hbox{$[7.7]$}}
\newcommand{\mirirten}{\hbox{$[10]$}}
\newcommand{\ab}{{\rm AB}}
\newcommand{\vega}{{\rm Vega}}
\newcommand{\ks}{\hbox{$K_s$}}
\newcommand{\tfit}{\hbox{TFIT}}
\newcommand{\sigmarms}{\sigma_\mathrm{rms}}
\newcommand{\gsim}{\gtrsim}
\newcommand{\lsim}{\lesssim}
\newcommand{\etal}{et al.}
\newcommand{\eg}{e.g.}
\newcommand{\ie}{i.e.}
\newcommand{\hbeta}{\hbox{H$\beta$}}
\newcommand{\he}{\hbox{H$\epsilon$}}
\newcommand{\hd}{\hbox{H$\delta$}}
\newcommand{\hg}{\hbox{H$\gamma$}}
\newcommand{\hc}{\hbox{H$\gamma$}}
\newcommand{\hb}{\hbox{H$\beta$}}
\newcommand{\ha}{\hbox{H$\alpha$}}
\newcommand{\oiii}{\hbox{O\,{\sc iii}}}
\newcommand{\oiiiuv}{\hbox{\ion{O}{3}]}}
\newcommand{\civ}{\hbox{\ion{C}{4}}}
\newcommand{\ciii}{\hbox{\ion{C}{3}]}}
\newcommand{\heii}{\hbox{\ion{He}{2}}}
\newcommand{\hei}{\hbox{\ion{He}{1}}}
\newcommand{\aliii}{\hbox{\ion{Al}{3}}}
\newcommand{\siiv}{\hbox{\ion{Si}{4}}}
\newcommand{\oiv}{\hbox{\ion{O}{4}]}}
\newcommand{\neiii}{\hbox{[\ion{Ne}{3}]}}
\newcommand{\neiv}{\hbox{[\ion{Ne}{4}]}}
\newcommand{\nv}{\hbox{[\ion{N}{5}]}}
\newcommand{\lstar}{\hbox{$L^\ast$}}
\newcommand{\extinctA}{\hbox{$A_{1700}$}}
\newcommand{\mathS}{\hbox{$\mathcal{S}$}}
\newcommand{\mathR}{\hbox{$\mathcal{R}$}}
\newcommand{\NLBG}{33} 
\newcommand{\dchi}{\hbox{$\Delta \chi^2$}}
\newcommand{\ebv}{\hbox{$E(B-V)$}}
\newcommand{\royalsoceity}{Phil.\ Trans.\ R.\ Soc.\ Lond.\ A}

\newcommand{\mstar}{\hbox{$M_\ast$}}
\newcommand{\Mstar}{\hbox{$M_\ast$}}

\newcommand{\zstar}{\hbox{$Z_\ast$}}
\newcommand{\Zstar}{\hbox{$Z_\ast$}}

\newcommand{\Msol}{\hbox{$M_\odot$}}
\newcommand{\Zsol}{\hbox{$Z_\odot$}}
\newcommand{\msol}{\hbox{$M_\odot$}}
\newcommand{\zsol}{\hbox{$Z_\odot$}}
\newcommand{\fion}[2]{\hbox{[\ion#1#2]}}
\newcommand{\infinity}{\hbox{$\infty$}}
\newcommand{\efolding}{$e$--folding}
\newcommand{\lya}{Lyman~$\alpha$}
\newcommand{\qso}{{Q0122+0338}}
\newcommand{\fos}{{FOS}}
\newcommand{\za}{\hbox{$z_\mathrm{a}$}}
\newcommand{\ze}{\hbox{$z_\mathrm{e}$}}
\newcommand{\NH}{\hbox{$N(\mathrm{H})$}}
\newcommand{\phc}{\phm{:}}
\newcommand{\inprep}{\textit{in prep}}
\newcommand{\distf}{\hbox{$f(M,\dot M,t)$}}
\newcommand{\degree}{\hbox{$^\circ$}}
\newcommand{\mydot}{\hbox{$\bullet$}}
\newcommand\myRoman[1]{\@Roman{#1}\relax}%
\newcommand{\fourge}{{\sc FourGE}}
\newcommand{\sfrten}{\hbox{SFR$_\mathrm{10}$}}
\newcommand{\sfrcen}{\hbox{SFR$_\mathrm{100}$}}

\newcommand{\mone}{\hbox{$[3.6]$}}
\newcommand{\mtwo}{\hbox{$[4.5]$}}
\renewcommand{\plotfiddle}[7]{
        \centering 
        \leavevmode
        \vbox to#2{\rule{0pt}{#2}}
        \includegraphics{#1}
}
\newcommand{\zfourge}{\hbox{ZFOURGE}}

\newcommand{\zftwo}{ZFK2}

\newcommand{\herschel}{\textit{Herschel}}
\newcommand{\jwst}{\textit{JWST}}
\newcommand{\kb}{\hbox{$K_b$}}
\newcommand{\kr}{\hbox{$K_r$}}
\newcommand{\jone}{\hbox{$J_1$}}
\newcommand{\jtwo}{\hbox{$J_2$}}
\newcommand{\jthree}{\hbox{$J_3$}}
\newcommand{\hs}{\hbox{$H_s$}}
\newcommand{\hl}{\hbox{$H_l$}}
\newcommand{\lir}{\hbox{$L_\mathrm{IR}$}}

\newcommand{\wfcj}{\hbox{$J_{125}$}}
\newcommand{\wfch}{\hbox{$H_{160}$}}
\newcommand{\OH}{\hbox{$12 + \log(\mathrm{O/H})$}}

\newcommand{\myhref}[1]{\href{#1}{#1}}

\newcommand{\lcdm}{$\Lambda$CDM}
\newcommand{\LCDM}{$\Lambda$CDM}
\renewcommand{\mone}{\hbox{$[3.6]$}}
\renewcommand{\mtwo}{\hbox{$[4.5]$}}
\newcommand{\mthree}{\hbox{$[5.8]$}}
\newcommand{\mfour}{\hbox{$[8.0]$}}
\newcommand{\reff}{\hbox{$r_\mathrm{eff}$}}
\newcommand{\sersic}{S\'ersic}
\newcommand{\um}{\hbox{$\mu$m}}
\newcommand{\pstart}[1]{\noindent \textbf{#1}}
\definecolor{aggiemaroon}{HTML}{500000}

\newcommand{\oii}{\hbox{[O\,{\sc ii}]}}
\newcommand{\nii}{\hbox{[N\,{\sc ii}]}}
\newcommand{\sii}{\hbox{[S\,{\sc ii}]}}
\newcommand{\hii}{\hbox{H\,{\sc ii}}}
\newcommand{\llambda}{\lambda\lambda}
\newcommand{\myshrink}{\vspace{-7pt}}
\newcommand{\mylinebreak}{\vspace{8pt}}

\newcommand{\grizli}{\hbox{\texttt{grizli}}}
\newcommand{\linmixi}{\hbox{\texttt{LINMIX}}}
\newcommand{\izi}{\hbox{\texttt{IZI}}}
\newcommand{\ppxf}{\hbox{\texttt{pPXF}}}
\newcommand{\bagpipes}{\hbox{\texttt{BAGPIPES}}}
\newcommand{\sextractor}{\hbox{\texttt{SE}}}

\newcommand{\fesc}{\hbox{$f_\mathrm{esc}$}}
\newcommand{\xiion}{\hbox{$\xi_\mathrm{ion}$}}
\newcommand{\luv}{\hbox{$L_\mathrm{UV}$}}
\newcommand{\muv}{\hbox{$M_\mathrm{UV}$}}
\newcommand{\ndotion}{\hbox{$\dot n_\mathrm{ion}$}}

\definecolor{aggiemaroon}{HTML}{500000}

\def\vshiftfig#1#2#3#4{\hfill 
\vbox{\parskip=0pt\hsize=#2
\vskip{#4} \includegraphics[width=#2]{#1}\vskip2pt
\vtop{\centering
\footnotesize
\hsize=#2
#3\vskip1pt
}}\hfill}